\documentclass[12pt,a4paper, fleqn]{article}
\usepackage[utf8]{inputenc}
\usepackage{longtable}
\usepackage{authblk}
\usepackage{geometry}
\usepackage{xcolor}
\usepackage{natbib}
\usepackage{graphicx}
\usepackage{pdflscape}  
\usepackage{appendix}
\usepackage{threeparttable}
\usepackage{booktabs}

\usepackage{amssymb}
\usepackage{bm}
\usepackage{amsmath}
\usepackage{graphicx}
\usepackage{mathtools}
\usepackage{rotating}
\usepackage{setspace}
\usepackage{color, colortbl}
\definecolor{ashgrey}{rgb}{0.7, 0.75, 0.71}
\definecolor{columbiablue}{rgb}{0.61, 0.87, 1.0}
\definecolor{coral}{rgb}{1.0, 0.5, 0.31}

\usepackage{enumitem}
\newlist{steps}{enumerate}{1}
\setlist[steps,1]{label = Step \arabic*:}

\usepackage{adjustbox}
\usepackage{dcolumn}
\newcolumntype{d}[1]{D..{#1}} 
\usepackage{diagbox, xcolor}

\usepackage{caption}
\usepackage{subcaption}
\definecolor{nblue}{HTML}{000660}
\usepackage{hyperref}

\usepackage{fancyhdr} 
\usepackage{tikz}
 
\usepackage{pifont}
\newcommand{\cmark}{\ding{51}}%
\newcommand{\xmark}{\ding{55}}%

\newcommand*{\myeqref}[2][Eq.~]{%
  \hyperref[{#2}]{#1(\ref*{#2})}%
}
\def\equationautorefname#1#2\null{%
  Eq.#1(#2\null)%
}

\begin{document}

\title{Bayesian Nonlinear Regression using Sums of  Simple Functions}
\author{Florian Huber\footnote{I would like to thank  Andrea Bastianin, Daniele Bianchi, Max Böck, Andrea Carriero, Liudas Giraitis, Fabrizio Iacone, Gary Koop, Massimiliano Marcellino, Michael Pfarrhofer, Michele Piffer and participants of the Econometrics Research Seminar at Queen Mary University of London and the Milan Time Series Seminar  for helpful comments and suggestions. Financial support by the Austrian Science Fund: ZK 35 is gratefully acknowledged.} \\ \textit{University of Salzburg}} 
\maketitle

\begin{abstract}
\noindent This paper proposes a new Bayesian machine learning model that can be applied to large  datasets arising in macroeconomics. Our framework sums over many simple two-component location mixtures. The transition between components is determined by a logistic function that depends on a single threshold variable and two hyperparameters.  Each of these individual models only accounts for a minor portion of the  variation in the endogenous variables. But many of them are capable of capturing arbitrary nonlinear conditional mean relations. Conjugate priors enable fast and efficient inference.  In simulations, we show that our approach produces accurate point and density forecasts. In a real-data exercise, we forecast US macroeconomic aggregates and consider the nonlinear effects of financial shocks in a large-scale nonlinear VAR.\\
\vspace{.15cm}
\end{abstract}


\normalsize\clearpage\doublespacing
\section{Introduction}
Nonlinear modeling of large datasets has received increasing attention in recent years. Extreme events such as the Covid-19 pandemic and the surge in inflation in the aftermath of the pandemic have raised the interest in more flexible econometric models \citep[see, e.g.,][]{Coulombe2020randomforest,  Coulombe2021covid, carriero2022addressing, hauzenberger2022fast, clark2023tail, huber2023nowcasting,koop2023bayesian}. 

Capturing nonlinearities in economic time series is predominantly achieved through estimating models with particular assumptions on the form of nonlinearities. For instance, Markov switching or structural break regressions and vector autoregressions  (VARs) assume that the parameters in the conditional mean change abruptly and there are only few but large breaks \citep[see, e.g.,][]{sims2006were, koop2007estimation, bauwens2015contribution}. By contrast, time-varying parameter (TVP) models \citep{primiceri2005time, COGLEY2005262, koop2013large, bitto2019achieving} assume that the parameters evolve smoothly over time and thus feature a large number of small breaks in the regression coefficients. 

All these methods have in common that they postulate a linear relationship between the endogenous variables and the regressors at  particular points in time. By contrast, nonlinear regression assumes a nonlinear relationship between the endogenous variables and the predictors. This relationship remains constant over time. Some examples are \cite{white1984nonlinear, hamilton2001parametric, lubrano2001smooth, hamilton2003oil, gerlach2008bayesian, gefang2009nonlinear, bruns2023tractable}. However, assuming a particular form of nonlinearity might give rise to model mis-specification and can be interpreted as a dogmatic Bayesian prior on the space of (nonlinear) conditional mean functions.

Another strand of the literature does not take a strong stance on the precise form of the conditional mean and uses nonparametric techniques to infer probable functional forms or detect structural breaks in the conditional mean. These methods remain agnostic on nonlinearities in the conditional mean and variances and try to infer them from the data. In recent years, nonparametric techniques have been increasingly used to forecast macro and financial aggregates \citep{clark2023tail, huber2023nowcasting}, estimate nonlinearities in key macroeconomic relations such as the Phillips curve \citep{Coulombe2020randomforest}, flexibly combine forecasts \citep{bassetti2018bayesian},  to construct shrinkage priors for vector autoregressions \citep[VARs, see][]{billio2019bayesian} and for pooling coefficients \citep{casarin2023bayesian}. The key shortcomings of these methods is that they are difficult to implement, customize and to tune.

These techniques all have their own pros and cons. However, what they share is the lack of scalability to very high dimensions. While there has been much progress in recent years \citep[see, e.g.,][]{chan2023large} the largest nonlinear models often feature less than 20 endogenous variables.  These 20 indicators often represent only a small fraction of the series available in different macroeconomic databases provided by  major central banks such as the US Federal Reserve or the Bank of England. For forecasting and structural analysis, exploiting as much information as possible can be important, increasing the demand for flexible models that can handle large datasets.

The last two paragraphs provide the main motivation for the current paper. We wish to develop techniques that are relatively simple to implement, modify and have the ability to handle large datasets commonly used in macroeconomics. These characteristics, however, should not come at the cost of reduced  flexibility.  We achieve this through a new parametric Bayesian nonlinear regression model that can be applied to univariate and multivariate time series and is inherently related to popular methods such as Bayesian additive regression trees \citep[BART, see][]{chipman2010bart} and shallow neural networks. Our main assumption is that the conditional mean is modeled  through a sum of simple functions. These functions are two-component location mixtures with transition between regimes  driven by a logistic transition function. The logistic function  is parameterized by a speed of adjustment coefficient, a threshold variable and a threshold parameter. These are all estimated through Bayesian techniques. When viewed individually, each of these simple models explains only a small fraction of the variation in the response (i.e., it acts as a 'weak learner'). However, when we sum over a moderate to large number of logistic functions we obtain a great deal of representation flexibility and end up with a model that is straightforward to estimate and to implement. 

The logistic function, while being tightly parameterized, is also flexible. For instance, if the speed of adjustment parameter becomes large, the transition function reduces to the indicator function that equals one if the threshold variable exceeds a threshold. If this applies for each of the individual functions we end up with an extreme version of BART with very simple trees.

Computation is carried out under conjugate priors. These provide further regularization but, more importantly, give rise to substantial computational gains. In particular, the algorithms we develop are highly scalable and can handle systems with hundreds of endogenous variables, leading to a huge dimensional nonlinear VAR model.

We start by illustrating our techniques by means of simulated data. Using a highly nonlinear DGP, we show that our parametric Bayesian model produces point and density forecasts that are often better than the ones produced by BART. We find that, as opposed to BART, the optimal number of functions to sum over is between 5 and 15 and thus much smaller. Moreover, we also find that fixing the speed of adjustment parameter so that the transition between regimes is instantaneous  yields  results that are only slightly worse than the ones from the model that estimates all parameters of the transition function.

We then move on to the real data analysis and estimate a large nonlinear VAR of the US economy.  This analysis consists of two parts. In the first, we show that our approach yields highly competitive density forecasts relative to the BART-VAR of \cite{clark2023tail}. In the second, we illustrate how our model can be used to analyze the nonlinear effects of financial shocks on the US economy. This exercise shows that for a dataset comprising of 80 endogenous variables, substantial asymmetries arise between benign and adverse shock. But this only holds if the shock is sufficiently large.

Our plan for the remainder of the paper is the following. We will introduce our main techniques in the next section. In this section, our focus is on approximating a nonlinear univariate regression model using sums of simple logistic functions. We provide an illustrating example, derive the likelihood, specify conjugate Bayesian priors for the parameters of the model and discuss posterior simulation. The next section, \autoref{sec: simulation}, provides simulation evidence for this model. Then, in \autoref{sec: aSTVAR}, we generalize the model to the multivariate case. \autoref{sec: IRFs} applies the model to a large US dataset and includes a forecasting exercise and the structural application. The final section summarizes and concludes the paper. 

\section{Parametric approximation to nonlinear regression}\label{sec:implications_model}
\subsection{The additive smooth transition regression} \label{ssec: addSTAR}
We start our discussion by focusing on the univariate case. Suppose that we have a time series $\{ y_t \}_{t=1}^T$ and  model it as a nonlinear function of a large panel of $K$ predictors $\bm x_t = (x_{1,t}, \dots, x_{K,t})' \in \mathbb{R}^K$. We approximate this nonlinear function using a sum of $J$ simpler functions (also called base learners):
\begin{equation}
    y_t = \sum_{j=1}^J g(\Tilde{x}_{j,t}|\bm \theta_j) + \varepsilon_t, \quad \varepsilon_t \sim \mathcal{N}(0, \sigma^2), \label{eq: nonlinear_reg}
\end{equation}
where $g: \mathbb{R} \to \mathbb{R}$ is a simple function that is fully parameterized by a low-dimensional vector $\bm \theta_j$. We will assume that $g$ is given by:
\begin{equation}
    g(\Tilde{x}_{j,t}|\bm \theta_j) = S_{j,t}(\Tilde{x}_{j,t}) \beta_{0,j} + [1 - S_{j,t}(\Tilde{x}_{j,t})] \beta_{1,j}, \label{eq: base learner}
\end{equation}
with $\beta_{i,j}~(i=0, 1)$ denoting a switching intercept term and $S_{j,t} \in [0, 1]$ is a transition function.  We let $\Tilde{x}_{j,t} = \bm \delta'_j \bm x_t$ denote an element of $\bm x_t$ and $\bm \delta_j$ is a $K-$dimensional selection vector. If the $s^{th}$ element of $\bm \delta_j$ equals $1$, the $s^{th}$ variable in $\bm x_t$ is selected and hence $\Tilde{x}_{j,t} = x_{s,t}$. In what follows, we suppress the dependence of $S_{j, t}$ on $\Tilde{x}_{j,t}$.

It is worth stressing that, as opposed to other algorithms, we only assume that a single variable informs the transition between two regimes. Using the jargon of the boosting literature \citep[for a survey, see][]{schapire2003boosting}, the function in \autoref{eq: base learner} acts as a weak learner \citep{bai2009boosting} and is expected to explain only a small amount of the variation in $y_t$. However, summing over multiple functions will provide sufficient representation flexibility to approximate any conditional mean function. This finding builds on theoretical results in \cite{cybenko1989approximation} and is closely related to the universal approximation theorem in the literature on machine learning.

For Theorem 1 in \cite{cybenko1989approximation} to work we need to make a few additional assumptions on the transition function $S_{j,t}$. In particular, we need to assume that $S_{j,t} = 0$ if $(\Tilde{x}_{j,t} \to -\infty$ and $S_{j,t} = 1$ if $(\Tilde{x}_{j,t} \to \infty$. A general function that fulfills this is the logistic function:
\begin{equation}
    S_{j,t} = \frac{1}{1 + \exp\{- \nu_j (\Tilde{x}_{j,t}-\mu_j)\}}, \label{eq: logistic}
\end{equation}
whereby $\nu_j \in \mathbb{R}^+$ is a speed of adjustment parameter and $\mu_j \in \mathbb{R}$ is a threshold parameter. The parameter $\nu_j$ controls the smoothness of the transition function. If it equals $0$, $S_{j,t}$ equals 1/2 and $\Tilde{x}_{j,t}$ does not enter $S_{j,t}$. If it is greater than zero but not too large we have a smooth transition between regimes with the transition being driven by the movements in $\Tilde{x}_{j,t}$. In this case, we would end up observing an S-shaped function. By contrast, if $\nu_j$ becomes large, we end up with an indicator function that equals zero if $\Tilde{x}_{j,t} > \mu_j$ and one otherwise. We call this model additive smooth transition  (AST) model and, for later convenience, we let $\bm \theta_j = (\mu_j, \nu_j, \bm \delta_j, \beta_{0, j}, \beta_{1, j})'$ denote the vector of component-specific parameters.

The main advantage of \autoref{eq: logistic} is that if $\Tilde{x}_{j,t}$ exerts a smooth effect (implying a gradual transition between regimes), the logistic function captures this through  estimates of $\nu_j$ closer to zero. By contrast, if $\Tilde{x}_{j,t}$ might only have a threshold effect, the model would estimate $\nu_j$ to be large and thus lead to a heavy side function.  By summing over many of these functions and allowing for the different parameters (thresholds and speed of adjustment coefficients) to vary our model provides a great deal of flexibility.

Our model is related to, at least, two popular models in the literature: BART and neural networks (NNs). BART is obtained if $g$ is replaced with a considerably more complex tree function. In this case, the dimension of the parameter vector $\bm \theta_j$ is not known a priori, rendering the model nonparametric. In many applications in a vast range of different fields, BART has been among the best performing specifications in terms of achieving low out-of-sample forecast errors \citep[see][]{chipman2010bart}. However, as opposed to our approach, if one wishes to apply BART to customized models (such as VARs) substantial coding efforts are required and while estimation of larger models is possible,\footnote{The largest BART model \cite{clark2023tail} consider features around 20 endogenous variables.} scalability to large simultaneous equation models such as the one we consider in our applied work, is currently unfeasible.

Another model closely related to the one presented in this section is the (shallow) NN. A shallow NN sets $\bm \delta_j \in \mathbb{R}^K$ equal to a weight vector. By doing so, every element in $\bm x_t$ informs the corresponding component-specific function. In addition, the transition functions often take different forms and enter the conditional mean equation as transformed regressors with separate coefficients. The key disadvantage relative to our approach is that it requires estimating a (possibly huge dimensional) coefficient vector per component function $J$. If $J$ becomes large, this becomes computational intensive and fully Bayesian inference is difficult to carry out in large models.

\subsection{Illustrating the mechanism}
Our model is best understood by considering a simple illustrative example where we fix $\nu_j$ and $\mu_j$. In this case, the intercept parameters $\beta_{0,j}$ and $\beta_{1,j}$ can be obtained through OLS. We consider the quarterly growth rate of US industrial production (IP) from $1990$:Q$1$ to $2019$:Q$4$. Our goal is to model IP growth as an unknown function of lagged IP growth and the excess bond premium (EBP) of \cite{gilchrist2012credit}. 

Consider the case of $J=1$ first and, let us assume, that $\Tilde{x}_{1,t}$ is the EBP, the threshold $\mu_1$ is the mean of the EBP and the speed of adjustment parameter is $\nu_1=0.3$, implying a smooth transition between regimes. The parameters $\beta_{0,1}$ and $\beta_{1,1}$ are estimated to be $-0.9$ and $1.4$, respectively. 

\begin{figure}[t]
    \centering
    \includegraphics[scale=0.47]{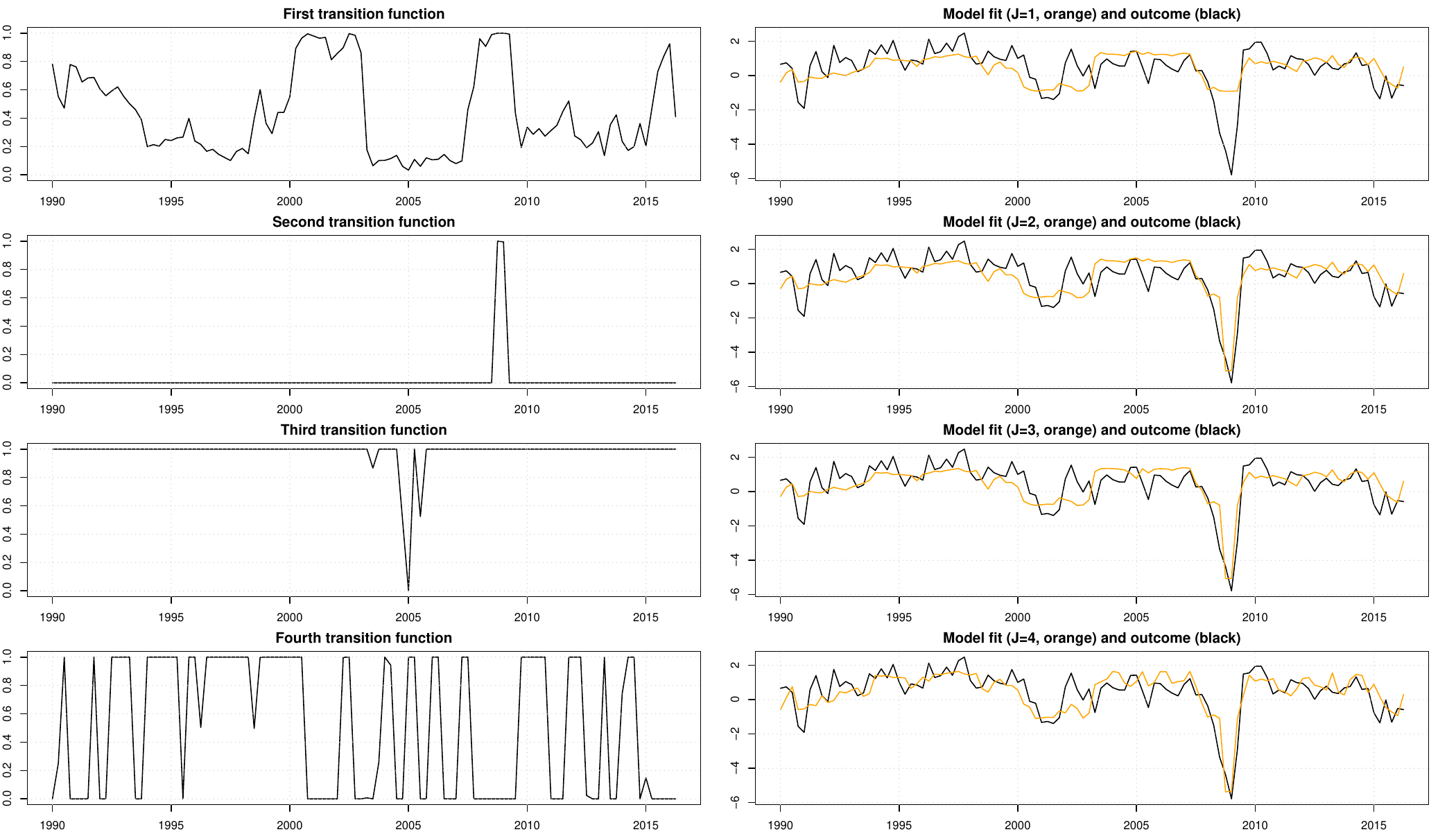}
    \caption{Transition functions and model fit for different values of $J$}
    \label{fig:illustr}
\end{figure}

The resulting transition function $S_{1,t}$ and fitted values are depicted in \autoref{fig:illustr}. Starting from top of the figure shows, in the left panel, the transition function $S_{1,t}$. Comparing the transition function with the outcome (right panel, black line) reveals that $S_{1,t}$ becomes large (approaches $1$) if IP growth is (strongly) negative. When we consider the fitted values, defined as:
\begin{equation*}
    \mathbb{E}(y_t|S_{1,t}) = -0.9 \times S_{1,t} + 1.4 \times (1-S_{1,t}),
\end{equation*}
we find that the first function already captures a considerable amount of variation in $y_t$. In particular, it succeeds in matching the slowly evolving local trends in IP growth. But it fails to capture much of the idiosyncratic behavior and, in particular, the substantial decline in IP growth during the 2008/2009 global financial crisis (GFC).

Consider adding a second component function. In this case, we fix $\nu_2=10$ and let $\mu_2$ be equal to the $0.99$ quantile of the EBP. With these parameter values, the transition function reduces to the indicator function that equals one if the EBP exceeds its $99$ percent quantile. Considering the transition function reveals that this is only the case during the GFC. In all other periods, the corresponding transition function is (almost) equal to zero with estimated parameters $\beta_{0, 2}= - 4.2$ and $\beta_{1, 2}=0.1$. Since our model is additive this implies that during the GFC, the growth rate is shifted downwards to reach approximately $-5.1$ percent.  Notice that a model with $J=2$ component functions is already capable of learning a great deal of variation in IP growth.

Increasing $J$ beyond two further improves the fit, but only slightly so. Using $J=3$ or $J=4$ (with transition functions being informed by the EBP in the case of $J=3$ and lagged IP growth for $J=4$) indicates that the estimated model fit displays more high frequency variation (in consistence with the actual time series). The key question, empirically, however is whether capturing more high frequency noise pays off for predictive performance. In our simulation study, we will return to this question and analyze the relationship between $J$ and predictive performance in more detail.

To sum up, in this simple toy example we find that summing over two logistic functions already provides a decent model fit. The first function, which is a smooth logistic function, explains low frequency trends whereas the second function captures the abrupt downturn during the GFC. 

\subsection{The likelihood}
Next we define the likelihood function of our model. To simplify the exposition, we let $\bm Z_t$ denote a $2J$-dimensional vector of (generated) regressors so that:
\begin{equation}
    \bm Z_t = (S_{1,t}, 1-S_{1,t}, \dots, S_{J,t}, 1-S_{J,t})'. \label{eq: Zs}
\end{equation}
In this case, we can rewrite \autoref{eq: nonlinear_reg} as follows:
\begin{equation*}
    y_t =  \bm \beta' \bm Z_t + \varepsilon_t,
\end{equation*}
with $\bm \beta = (\beta_{0, 1}, \beta_{1,1}, \dots, \beta_{0, J}, \beta_{1, J})'$ being a vector of stacked coefficients.  Stacking over $t$ gives rise to the full-data representation of the model:
\begin{equation}
    \bm y = \bm Z \bm \beta + \bm \varepsilon, \label{eq: stacked_nonlinear_reg}
\end{equation}
where $\bm y =(y_1, \dots, y_T)', \bm Z = (\bm Z_1, \dots, \bm Z_T)'$ and $\bm \varepsilon = (\varepsilon_1, \dots, \varepsilon_T)'$ are $T \times 1$, $T \times 2J$ and $T \times 1$ matrices, respectively.

Standard textbook results \citep[see, e.g,][]{chan2019bayesian} show that the likelihood function can be rewritten as:
\begin{equation}
    p(\bm y|\bm Z, \bm \beta, \sigma^{-2}) \propto (\sigma^2)^{2J} \exp \left[  - \frac{1}{2 \sigma^2} \left( \bm \beta - \hat{\bm \beta}\right)' \bm Z' \bm Z \left(\bm \beta - \hat{\bm \beta}\right)\right] \times \left[ (\sigma^2)^{\frac{w}{2}} \exp      \left( - \frac{w}{2 \sigma^2 s^{-2}} \right)  \right]. \label{eq: likelihood}
\end{equation}
Here, we let $\hat{\bm \beta} = (\bm Z' \bm Z)' \bm Z' \bm y$ denote the OLS/maximum likelihood estimator of $\bm \beta$, $w = T - 2J$ the degrees of freedom, and $s^2 = \frac{(\bm y - \bm Z \hat{\bm \beta})' (\bm y - \bm Z \hat{\bm \beta})}{w}$ is the OLS estimator of the error variance. Notice that since $\bm Z$ depends on the speed of adjustment parameters, thresholds and threshold variables, we do not condition on $\{\nu_j\}, \{\mu_j\}$ and $\{ \bm \delta_j \}$. The likelihood function consists of two terms.  The first term implies a dependence between $\bm \beta$ and $\sigma^2$ whereas the second term is independent of $\bm \beta$ and looks like the kernel of an inverse Gamma distribution. We will use these observations to construct a standard conjugate prior in the next section. 

\subsection{The prior}
The model in \autoref{eq: nonlinear_reg} might be subject to overfitting if $J$ is set too large. Hence, we need to regularize the estimates of $\bm \beta$. This is achieved through shrinkage priors that are inspired by the priors stipulated in \cite{chipman2010bart}. Our joint prior on the parameters of the model can be factorized as follows:
\begin{equation}
    p(\bm \beta, \sigma_2, \{\mu_j\}, \{\nu_j\}, \{\delta_j\}) = p(\bm \beta|\sigma^2)~p(\sigma^2)~\prod_{j=1}^J\left( p(\mu_j) ~ p(\nu_j) ~p(\bm \delta_j) \right). \label{eq: jointprior}
\end{equation}
Note that the prior on $\bm \beta$ depends on $\sigma^2$ while the priors on the other parameters are independent of each other.   We assume that $p(\bm \beta|\sigma^2)$ is Gaussian:
\begin{equation}
    p(\bm \beta|\sigma^2) = \mathcal{N}(\bm 0, \sigma^2 \underline{\bm V}),
\end{equation}
where $\underline{\bm V} = \phi J^{-1} \times \bm I_{2J}$ and $\phi$ is a positive prior scaling parameter. The prior variance decreases in $J$ and hence, for a large number of component functions, we shrink the parameters stronger to zero so that each function is expected to contribute less to explain the variation in $y_t$.  Give that $\bm Z_t$ is bounded between $0$ and $1$, we set $\phi=1$.

On the error variances we use the usual inverse Gamma prior $p(\sigma^2) = \mathcal{G}^{-1}(\underline{a}_\sigma, \underline{b}_\sigma)$. We let $\underline{a}_\sigma$ and $\underline{b}_\sigma$ denote the prior degree of freedom and a prior scaling parameter, respectively. To render this prior effectively uninformative, we set $\underline{a}_\sigma = \underline{b}_\sigma = 0.01$. Notice that one could also use a data-driven prior that is scaled by, e.g., the OLS standard deviation or other estimates of the error variances. If this estimate implies under-dispersion one could then place more weight on the prior to shrink the error variances towards zero and thus force the conditional mean to soak up more variation in $y_t$. 

For the thresholds we use weakly informative Gaussian priors $p(\mu_j) = \mathcal{N}(0, \underline{\sigma}^2_j)$ where $\underline{\sigma}^2_j$ is a hyperparameter which we set to a large value.  In our case, we standardize the input data by subtracting the mean and normalizing by the standard deviation. Hence, the prior centers the threshold over the mean of the non-normalized version of $\Tilde{x}_{j, t}$.  Using more informative priors on the thresholds is difficult, in particular given the fact that we do not consider a standard smooth transition model where prior information about possible thresholds could exist. 

On $\nu_j$ we use an weakly informative inverse Gamma prior $p(\nu_j) = \mathcal{G}^{-1}(\underline{a}_\nu, \underline{b}_\nu)$ where the hyperparameters $\underline{a}_\nu =\underline{b}_\nu =0.01$ are set close to zero throughout the paper. This choice has been used in, e.g.,  \cite{lopes2006bayesian}.\footnote{\cite{lubrano2001smooth} discusses an alternative based on a truncated Cauchy prior on $\nu_j$. This choice would be straightforward to adopt in our setting.} Finally, we use a discrete uniform prior on $\bm \delta_j$ so that each element is equally likely to be equal to zero.

\subsection{Posterior simulation}\label{ssec: posterior}
The prior in \autoref{eq: jointprior} can be combined with the likelihood in \autoref{eq: likelihood} to derive the joint posterior:
\begin{align*}
    p(\bm \beta, \sigma^2, \{\mu_j\}, \{\nu_j\}, \{\delta_j\} | \bm y) = p(\bm y|\bm Z, \bm \beta, \sigma^2) \times  p(\bm \beta|\sigma^2)~p(\sigma^2)~\prod_{j=1}^J\left( p(\mu_j) ~ p(\nu_j) ~p(\bm \delta_j) \right).
\end{align*}
This joint posterior takes no well known form. However, given that the prior on $\bm \beta$ and $\sigma^2$ are conditionally (on $\bm Z$) conjugate, we can make use of the fact that $p(\bm \beta, \sigma^2|\bm y, \bm Z)$ takes a well known form \citep[see, e.g.,][]{koop2003bayesian}:
\begin{equation*}
    p(\bm \beta, \sigma^2|\bm y, \bm Z) = \text{NIG}(\overline{\bm \beta}, \overline{\bm V}_\beta, \overline{a}_\sigma, \overline{s}_\sigma),
\end{equation*}
where NIG denotes the Normal-Inverse Gamma distribution with four parameters:
\begin{equation*}
   \overline{\bm \beta} = \overline{\bm V}_\beta (\bm Z \bm y), \quad \overline{\bm V}_\beta = (\underline{\bm V}_\beta + \bm Z' \bm Z)^{-1}, \quad
   \overline{a}_\sigma = \underline{a}_\sigma + T, \quad \overline{s}_\sigma =  \underline{b}_\sigma + \frac{1}{2}\left(\overline{\bm \beta}' \overline{\bm V}^{-1}_\beta \overline{\bm \beta}\right).
\end{equation*}
The parameters associated with the transition functions are then sampled using the Bayesian backfitting strategy outlined in \cite{hastie2000bayesian} and used in \cite{chipman2010bart}. We let:
\begin{equation*}
    R_{jt} = y_t - \sum_{s \neq j} g(\bm \Tilde{x}_{s, t}|\bm \theta_s)
\end{equation*}
denote the partial residual vector and $\bm R_j = (R_{j1}, \dots, R_{jT})'$. Moreover, let $\bm Z_j$ denote a $T \times 2$ matrix with $t^{th}$ row equal to  $(S_{jt}, (1-S_{jt}))$ and $\bm \beta_j = (\beta_{0, j}, \beta_{1, j})'$. 

The conjugacy of our prior setup implies that we can integrate out $\bm \beta_j$ and $\sigma^2$ to obtain:
\begin{equation}
    p(\bm R_j | \nu_j, \mu_j, \bm \delta_j) \propto \sqrt{\frac{|\overline{\bm V}_j|}{|\underline{\bm V}_j|}} \times \left[\underline{b}_\sigma + \left(\bm R'_j \bm R_j - \frac{1}{2} \overline{\bm \beta}'_j \overline{\bm V}^{-1}_j \overline{\bm \beta}_j\right)  \right]^{-\frac{\underline{a}_\sigma + T}{2}}. \label{eq: ML_R}
\end{equation}
Here, we let $\overline{\bm \beta}'_j =\overline{\bm V}_j \bm Z_j \bm R_j$ and $\overline{\bm V}_j = (\bm Z'_j\bm Z_j + J \phi^{-1} \bm I_2)^{-1}$ denote the posterior mean and variance of $\bm \beta_j$, respectively. Notice that we implicitly condition on the other $\bm \beta_s$ and $\bm Z_s$ for $s \neq j$. 

We then sample $p(\nu_j, \mu_j, \bm \delta_j|\bm R_j) \propto p(\bm R_j | \nu_j, \mu_j, \bm \delta_j) \times p(\nu_j, \mu_j, \bm \delta_j)$ in two blocks. First, we let $\Tilde{\delta}_j \in \{1, \dots, K\}$ denote a categorical auxiliary variable that indicates the element in $\bm \delta_j$ which equals one. The posterior probability that  $\Tilde{\delta}_j = i$ is then given by:
\begin{equation}
    \text{Prob}(\Tilde{\delta}_j = i|\nu_j, \mu_j, \bm R_j) \propto p(\bm R_j | \tilde{\delta}_j = i, \nu_j, \mu_j) \times p(\bm \delta_j), \label{eq: probs_delta}
\end{equation}
and we can easily compute \autoref{eq: probs_delta} for all $j=1, \dots, K$. 

Conditional on $\bm \delta_j$ we sample $\nu_j$ and $\mu_j$ jointly using a single random walk Metropolis Hastings step where we propose $(\nu^*_j, \mu^*_j)' \sim \mathcal{N}((\nu^a_j, \mu^a_j)', \text{diag}(s_\nu, s_\mu))$, with the superscript $a$ indicating the previous accepted draw. The scaling parameters of the proposal distribution are tuned during the first half of the burn-in stage of our algorithm so that the acceptance probability is between 30 and 60 percent. After proposing $\nu^*_j, \mu^*_j$, we accept the proposed values with probability equal to:
\begin{equation*}
    \alpha((\nu^*_j, \mu^*_j), (\nu^a_j, \mu^a_j)) = \text{min}\left( \frac{p(\bm R_j|\nu_j = \nu_j^*, \mu_j = \mu_j^*, \delta_j) \times p(\nu_j = \nu_j^*, \mu_j = \mu_j^*)}{p(\bm R_j|\nu_j = \nu^a_j, \mu_j = \mu^a_j, \delta_j) \times p(\nu_j=\nu_j^a, \mu_j= \mu_j ^a)}, 1 \right).
\end{equation*}

This completes the different steps to sample from the relevant full conditional posterior distributions. Since we sample some parameters marginal of the others, the ordering of the steps of the sampler play an important role \citep{van2008partially}. Taking this into account, our algorithm cycles between the following steps:
\begin{enumerate}
    \item Sample $\tilde{\delta}_j|\bm R_j \sim p(\tilde{\delta}_j|\nu_j, \mu_j, \bm R_j)$ using \autoref{eq: probs_delta}.
    \item Sample $\nu_j$ and $\mu_j$ in a block using the MH updating step outlined above.
    \item Sample the error variances  $\sigma^2|\bm Z \sim \mathcal{G}^{-1}(\overline{a}_\sigma, \overline{s}_\sigma)$ from an inverse Gamma distribution.
    \item Sample $\bm \beta|\sigma^2, \bm Z \sim \mathcal{N}(\overline{\bm \beta}, \sigma^2 \overline{\bm V}_\beta)$ from a Gaussian conditional posterior distribution given $\bm Z$ and $\sigma^2$.
\end{enumerate}
The first two steps are marginal of $\bm \beta$ and $\sigma^2$ while Step 3 is conditional on $\bm Z$ (i.e. $\{\nu_j, \mu_j, \bm \delta_j\}$) but marginal of $\bm \beta$. The final step is conditional on the error variances and the different component functions. The key property of this algorithm is that we exploit conjugacy to sample the parameters of the component functions independently from the error variances and the regression coefficients. This improves mixing substantially and we found that in our empirical work and the simulations that our algorithm converges rapidly towards the desired stationary distribution. 

\section{Monte Carlo evidence}\label{sec: simulation}
In this section we put our proposed model to a test within a controlled environment. In particular,  we show that under a nonlinear data generating process (DGP), our model yields predictions which are accurate and can compete with the ones of BART. The reason why we benchmark the results to BART is due to the empirical success of BART across many fields. We use a standard BART implementation with precisely the same set of hyperparameters on the trees and the error variances as in \cite{chipman2010bart}.

We assume that $\{y_t\}_{t=1}^{T=300}$ is generated as follows:
\begin{equation*}
   y_t = 0.9 y_{t-1} + \bm \beta_{\text{true}} \bm x_{t-1} +\bm \kappa_{\text{true}}  \bm x^2_{t-1} + u_t, \quad u_t \sim \mathcal{N}(0, 1)
\end{equation*}
where $\bm x_t \sim \mathcal{N}(\bm 0, \bm I)$ is a $K=25$-dimensional vector, $ \bm \beta_{\text{true}} \sim \mathcal{N}(3, 9)$ is a $25$-dimensional vector of true linear coefficients and $\bm \kappa_{\text{true}} \sim \mathcal{N}(2, 9)$ is a $K-$dimensional vector of nonlinear coefficients. To  have a sparse model we zero out 60\% of the elements in both $\bm \beta_{\text{true}}$ and $\bm \kappa_{\text{true}}$. Finally, we initialize $y_0 = 0$. This DGP produces time series that match patterns commonly observed in macroeconomics and finance. 



We estimate four variants of the AST model for different values of $J$. The first estimates $\nu_j$ and $\mu_j$ using the prior setup discussed in the previous section. The second fixes $\nu_j = 10$, leading to a model that sums over mixtures connected by a threshold function. The third assumes $\mu_j = \hat{\mu}_j = \sum_{t=1}^T \bm \delta_j \bm x_t/T$, implying that the threshold is the empirical mean of the corresponding variable selected by $\bm \delta_j$. Finally, the last specification fixes $\nu_j = 1$ and $\mu_j = \hat{\mu}_j$, leading to a model which sets $S_{jt} = 1$ if $\tilde{x}_{j, t}$ exceeds its mean. The first model is the most flexible one and allows for different threshold values and different speed of adjustments of the transition functions. The last one introduces strong restrictions. The intermediate specifications provide slightly more flexibility and by doing so reduce the number of free parameters. In our simulation we investigate how these choices impact the point and density forecasting performance. To simulate a high dimensional setting, we include four lags of the regressors.

We carry out our forecasting exercise by taking each generated series $\{y_t\}$ and splitting it into two halves of equal size. The first half $t=1, \dots, T_0 ( = T/2 = 150)$ is used to train each model whereas we predict the second half $T_0+1 (=151), \dots, T (=300)$. To speed up computation and due to the fact that our DGP features no structural breaks, we only estimate the model once and then compute one-step-ahead predictions. To control for sampling uncertainty with respect to the DGP we repeat these experiments $50$ times.

To analyze forecast accuracy we compute the root mean squared error (RMSE) as follows:
\begin{equation*}
    \text{RMSE} = \sqrt{\frac{1}{T-T_0} \left( \sum_{t=T_0+1}^{T}(y_{t} -  \overline{y}_{t|t-1})^2\right)},
\end{equation*}
where $\overline{y}_{t|t-1}$ denotes the median of the one-step-ahead predictive density. To measure the accuracy of density forecasts we compute the log predictive likelihood (LPL) using a Gaussian approximation:
\begin{equation*}
    \text{LPL} = \frac{1}{T-T_0} \left( \sum_{t=T_0+1}^{T} \log \mathcal{N}(y_t |\overline{y}_{t|t-1}, \overline{\sigma}^2_{t|t-1})\right),
\end{equation*}
with $p(y_t |\overline{y}_{t|t-1}, \sigma^2_{t|t-1})$ being the predictive distribution evaluated at the actual outcome and  $\overline{\sigma}^2_{t|t-1}$ denoting the predictive variance.

Table \ref{tab: sim} shows the results of this simulation exercise. The upper panel of the table shows the RMSEs relative to BART so that numbers greater than one suggest a weaker point forecasting accuracy whereas number smaller than one point towards outperformance of a corresponding AST model.  The lower panel shows differences in LPLs between a given AST specification and the BART benchmark, with numbers greater than zero suggesting more precise density forecasts and negative numbers point towards a weaker average density forecast performance.

Considering RMSE results reveals that our baseline specification that estimates $\nu_j$ and $\mu_j$ yields forecasts that improve upon the BART forecasts for $J$ between five and $25$. The improvements in relative RMSEs are U-shaped and first increase until $J=10$, becoming smaller afterwards. At a first glance, this suggests that careful selection of $J$ is necessary to produce accurate forecasts. However, it is worth stressing that BART-based forecasts are typically very precise and our approach, being simpler to implement and, as we will see in the next sections, more scalable, never loses against BART as long as $J > 1$. 

If we consider the specification that fixes the threshold parameters, we find a  weaker overall performance but RMSE ratios are still $\pm 10$ percent within the absolute RMSEs of BART for all values of $J$. Similar results arise if we fix the speed of adjustment parameters but estimate thresholds. In this case, the $J=1$ case performs poorly. This is expected given that this model is a simple switching model with endogenous selected threshold variable and estimated threshold. If we increase the number of component functions the performance increases until $J=25$. Finally, not estimating $\nu_j$ nor $\mu_j$ is not a good idea. In this case, we lose against the BART benchmark by large margins.

\begin{table}[ht]
\centering
\begin{tabular}{llllrrrrrr}
  \toprule
 & \multicolumn{2}{c}{Est.} & & \multicolumn{6}{c}{$J=$}\\
 & $\nu$ & $\mu$ &  &   1 & 5 & 10 & 15 & 25 & 50 \\ 
  \hline
RMSE & \cmark & \cmark &  &  1.06 &  0.95 &  0.87 &  0.89 &  0.98 &  1.00 \\ 
    & \cmark & \xmark &  &  1.00 &  0.93 &  0.97 &  0.97 &  1.02 &  1.09 \\ 
 &\xmark  & \cmark &   &  1.55 &  1.09 &  0.97 &  0.91 &  0.91 &  1.05 \\ 
   & \xmark  & \xmark  &  &  2.36 &  2.12 &  1.91 &  1.96 &  1.80 &  1.79 \\ \midrule
 LPL & \cmark  & \cmark &  &  0.15 &  0.26 &  0.36 &  0.33 &  0.21 &  0.17 \\ 
   & \cmark & \xmark &  &  0.21 &  0.28 &  0.24 &  0.21 &  0.15 &  0.06 \\ 
   & \xmark & \cmark &  & -0.26 &  0.06 &  0.21 &  0.28 &  0.30 &  0.11 \\ 
   & \xmark &\xmark  &  & -0.70 & -0.58 & -0.49 & -0.48 & -0.45 & -0.44 \\ 
   \bottomrule
\end{tabular}
\caption*{\footnotesize \textbf{Notes:} \cmark and \xmark~ denote whether $\nu$ and/or $\mu$ is estimated or kept fixed, respectively. In case $\nu$ is fixed, we set it equal to $\nu=10$, implying that $S_{jt}$ is the indicator function. In case we fix $\mu$, we set it to $\mu=0$. This implies that the mean of the series is used as a threshold variable. Results are ratios to the BART RMSEs and differences to the BART LPLs, respectively.}
\caption{Simulation results}\label{tab: sim}
\end{table}

Next, we consider the density forecasting performance in the lower panel of \autoref{tab: sim}. Recall that numbers greater than zero indicate outperformance of AST whereas negative numbers suggest the opposite. In principle, the density forecasting results tell a story similar (but slightly more pronounced) to the RMSE results. Depending on the choice of $J$, AST improves upon BART and the most flexible version does best on average. The key difference, however, is that for the model that estimates $\nu_j$ and $\mu_j$ we find improvements in accuracy for all values of $J$. But, similar to the findings for point forecasts, these improvements first increase with $J$ and then slowly decay.  Out of the restricted versions we find that the model which estimates the speed of transition parameter performs best, yielding gains for different values of $J$. The one that performs worst is, again, the model that fixes both $\nu_j$ and $\mu_j$. 

Our previous discussion has established that AST yields forecasts which are often better than the ones produced by BART. A general conclusion stemming from our synthetic data exercise is that for the most flexible version, setting $J>1$ yields point forecasts which are, in the worst case, very close to the ones produced by BART and always produces slightly more accurate density predictions. Another question relevant for practitioners, however, is whether the model performs well in selecting the correct covariates. This is what we investigate in \autoref{tab: var_relevance} by looking at a particular realization from the DGP.

\begin{table}[t!]
\centering
\scalebox{0.86}{
\begin{tabular}{lrr|rrrr}
  \toprule
  \multicolumn{3}{c}{} & \multicolumn{4}{c}{Variable relevance}\\
 & $\bm \beta_{true}$ & $\bm \kappa_{true}$ & $p=1$ & $p=2$ & $p=3$ & $p=4$ \\ 
  \hline
$y_{t-1}$ & 0.90 & 0.00 & 3.82 & 0.08 & 0.07 & 0.06 \\ 
$x_{1, t}$ & 0.00 & 0.00 & 0.05 & 0.05 & 0.05 & 0.05 \\ 
$x_{2, t}$ & 5.83 & 0.00 & 1.05 & 0.05 & 0.05 & 0.06 \\ 
$x_{3, t}$ & 0.00 & 0.00 & 0.06 & 0.06 & 0.06 & 0.05 \\ 
$x_{4, t}$ & 0.00 & 0.65 & 0.05 & 0.05 & 0.05 & 0.05 \\ 
$x_{5, t}$ & 0.00 & 0.00 & 0.05 & 0.05 & 0.05 & 0.05 \\ 
$x_{6, t}$ & 0.00 & 3.59 & 0.07 & 0.05 & 0.05 & 0.05 \\ 
$x_{7, t}$ & 0.00 & 0.00 & 0.06 & 0.05 & 0.05 & 0.04 \\ 
$x_{8, t}$ & 0.00 & 0.00 & 0.05 & 0.05 & 0.05 & 0.05 \\ 
$x_{9, t}$ & -2.87 & 5.73 & 1.06 & 0.06 & 0.05 & 0.05 \\ 
$x_{10, t}$ & 0.00 & 4.57 & 0.51 & 0.05 & 0.06 & 0.05 \\ 
$x_{11, t}$ & 0.00 & 0.00 & 0.05 & 0.05 & 0.06 & 0.06 \\ 
$x_{12, t}$ & 2.58 & 0.37 & 0.05 & 0.05 & 0.05 & 0.06 \\ 
$x_{13, t}$ & 3.65 & 4.06 & 0.07 & 0.05 & 0.05 & 0.05 \\ 
$x_{14, t}$ & 0.00 & -0.43 & 0.07 & 0.05 & 0.05 & 0.06 \\ 
$x_{15, t}$ & 0.00 & 0.00 & 0.05 & 0.05 & 0.05 & 0.06 \\ 
$x_{16, t}$ & 0.00 & 0.00 & 0.05 & 0.08 & 0.05 & 0.05 \\ 
$x_{17, t}$ & 0.00 & 0.00 & 0.06 & 0.05 & 0.06 & 0.05 \\ 
$x_{18, t}$ & 0.00 & -1.42 & 0.06 & 0.05 & 0.05 & 0.06 \\ 
$x_{19, t}$ & 6.90 & 0.00 & 1.02 & 0.05 & 0.05 & 0.06 \\ 
$x_{20, t}$ & 8.20 & 0.00 & 1.07 & 0.05 & 0.07 & 0.05 \\ 
$x_{21, t}$ & 4.80 & 5.18 & 1.06 & 0.05 & 0.06 & 0.06 \\ 
$x_{22, t}$ & 3.14 & -0.22 & 0.19 & 0.05 & 0.06 & 0.06 \\ 
$x_{23, t}$ & -0.90 & 0.00 & 0.05 & 0.07 & 0.07 & 0.05 \\ 
$x_{24, t}$ & 4.94 & 0.00 & 0.07 & 0.06 & 0.06 & 0.05 \\ 
$x_{25, t}$ & 0.00 & 0.00 & 0.05 & 0.05 & 0.05 & 0.05 \\ 
   \bottomrule
\end{tabular}
}
 \caption*{\footnotesize \textbf{Notes:} The columns $'\beta_{true}'$ and $'\kappa_{true}'$ denote the true coefficients. The columns 'Variable relevance' denote the sum of the posterior means of the indicators $\delta_{jt}$ for all $j$ and across the different lags.}
 \caption{Variable relevance and true parameter values for a single realization from the DGP}\label{tab: var_relevance}
\end{table}

The first two columns of the table show the actual values of $\bm \beta_{\text{true}}$ and $\bm \kappa_{\text{true}}$ and the remaining columns show variable relevance scores for the different lags of $\bm x_t$. These are computed by taking the posterior mean of $\bm \delta_i, \overline{\bm \delta}_i$, and then summing over all $i$. A given number hence indicates how often a variable shows up in \textit{all} component functions and greater numbers thus imply a higher variable relevance. If a score is close to one it implies that only one of the base functions includes a given variable in $\bm x_t$.

At a very general level, we find a close association between regressors that feature large values of $\bm \beta_{true}$ and/or $\bm \kappa_{true}$. If this is the case, variable relevance scores are often above one. There are some rare cases where this does not hold (such as $x_{13,t}$ and $x_{14,t}$), but for the vast majority our model attributes appreciable relevance to covariates that feature large coefficients (in absolute terms).  Variables that do not enter the DGP are, without any exception, never included in the corresponding base learners and thus do not impact our model. The single most variable, as expected, is the first lag of the endogenous variable, which shows up almost four out of $J$ times in the corresponding functions.

\begin{figure}
    \centering
    \includegraphics[scale=0.33]{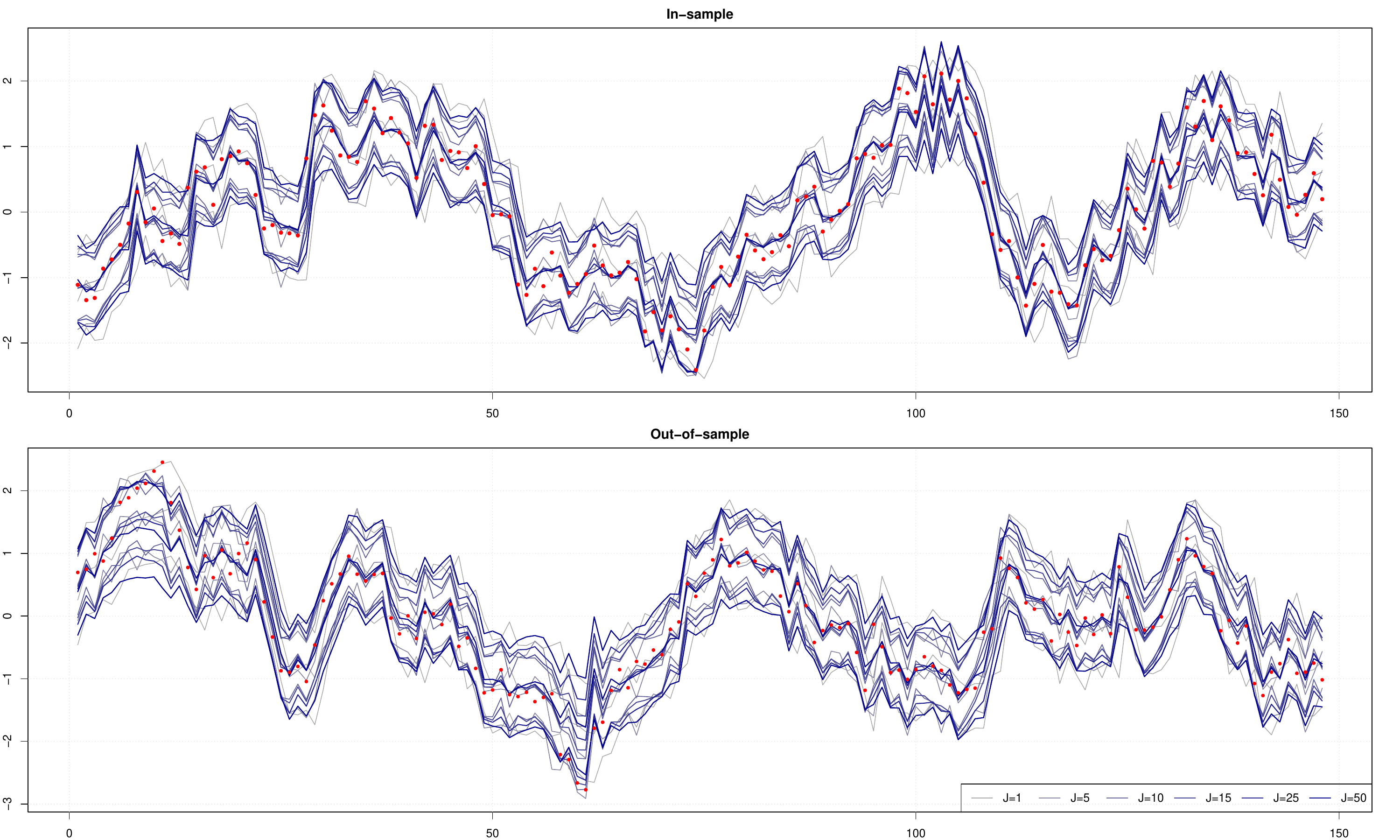}
    \caption{In-sample and out-of-sample densities for different values of $J$}
    \label{fig:pred_dens}
\end{figure}

To shed light on the differences in the predictive densities across different values of $J$, \autoref{fig:pred_dens} plots the in-sample and out-of-sample predictive densities for different values of $J$ for the baseline model that estimates both $\nu_j$ and $\mu_j$. In both cases, the results reveal that the model does a good job in fitting the data, irrespective of the choice of $J$. In principle, there are no discernible differences in terms of the posterior medians. The only feature that stands out is that credible sets  become smoother and slightly tighter for larger values of $J$ in-sample and, to a somewhat lesser degree, out-of-sample. 

Next, we investigate the relationship between $J$ and the shape of the transition functions. To this end, we compute the average transition function based on the (normalized) predictors. This is achieved as follows.  For each $j=1, \dots, J$,  we take the posterior median  of $\nu_j, \mu_j$ and $\bm \delta_j$ and plot the transition function for $\tilde{x}_{j, t}$ ranging from $-10$ to $10$. This gives rise to $J$ different transition functions. When then simply compute the average across these $J$ transition functions and plot these. Hence, the resulting average transition function reflects how the average base learner moves from $S_{j,t} = 0$ to  $S_{j, t}=1$.

\begin{figure}
    \centering
    \includegraphics[scale=0.43]{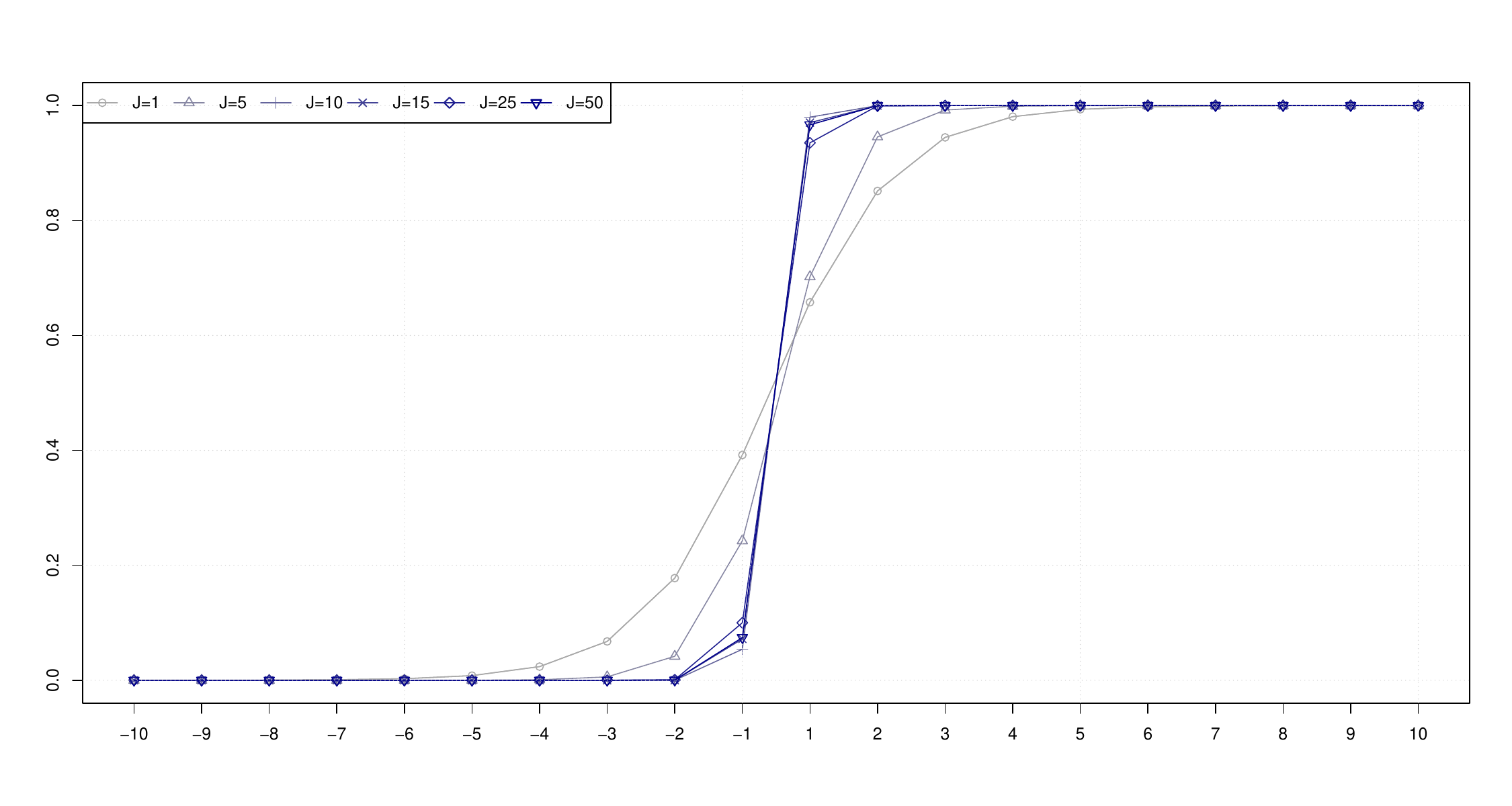}
    \caption{Estimated transition function. $\nu$ and $\mu$ averaged across the different submodels.}
    \label{fig:transition_average}
\end{figure}

\autoref{fig:transition_average} shows the shape of these average functions. The single most striking observation is that the speed of adjustment parameter seems to increase with $J$. Whereas we find a rather smooth transition for $J=1$ and $J=5$, going from $J=10$ to $J=15$ implies a transition that is very close to using an indicator function. This indicates that if we use only few base functions to learn the conditional mean relations, our algorithm places substantial posterior mass on transition functions that feature more complex patterns. But for larger $J$, the individual functions become simpler. It is, however, worth stressing that the  average transition function for $J=50$ is still not exactly equal to an indicator function and still implies a somewhat gradual transition between regimes for values of $\tilde{x}_{j,t}$ close to the mean. 

\section{The vector additive smooth transition model} \label{sec: aSTVAR}
\subsection{The likelihood}\label{subsec:NEset}
In the previous sections we have developed the AST model and illustrated its usefulness in Monte Carlo simulations. In this section, we generalize the model to the multivariate case and develop a scalable, conjugate version of it to model a possible large panel of $M$ macroeconomic time series which we store in $\bm y_t$.  This model is henceforth labeled the vector additive smooth transition (VAST) model.

We assume that $\bm y_t$ depends nonlinearily on its $P$ lags. These are stored in a $K$-dimensional vector $\bm x_t = (\bm y'_{t-1}, \dots, \bm y'_{t-P})'$ with $K = MP$. The vector additive smooth transition (VAST) model is then given by:
\begin{equation}
    \bm y_t = \sum_{j = 1}^J g(\tilde{x}_{j, t}, \bm \theta_j) + \bm \varepsilon_t, \quad \bm \varepsilon_t \sim \mathcal{N}(\bm 0_M, \bm \Sigma), \label{eq: VASTM}
\end{equation}
where $g: \mathbb{R} \to \mathbb{R}^M$ is a function that maps a scalar input $\tilde{x}_{j, t}$ into an $M$-dimensional output and $\bm \varepsilon_t$ is a Gaussian white noise process with zero mean and covariance matrix $\bm \Sigma$. 

The component function $g$ takes the following form:
\begin{equation}
    g(\tilde{x}_{j, t}|\bm \theta_j) = S_{j, t} \bm \beta_{0, j} + (1- S_{j, t}) \bm \beta_{1, j}. \label{eq: str}
\end{equation}
This transition function looks similar to \autoref{eq: base learner} but the location parameters $\bm \beta_{i, j} = (\beta_{ij, 1}, \dots, \beta_{ij, M})'$ are now $M$-dimensional vectors. We will again assume that $S_{j, t}$ takes precisely the same form as \autoref{eq: logistic}.  

Under Eqs. (\ref{eq: VASTM}) to (\ref{eq: str}), the model can be written as:
\begin{equation*}
    \bm y_t = (\bm I_M \otimes \bm Z'_t) \bm \beta + \bm \varepsilon_t,
\end{equation*}
where $\bm Z_t$ is given by \autoref{eq: Zs} and $\bm \beta = (\bm \beta'_{0, 1}, \bm \beta'_{1, 1}, \dots, \bm \beta'_{0, J}, \bm \beta'_{1, J'})'$ is a $N-$dimensional vector with $N=2JM$. 

The number of free parameters in the model is $v^{\text{VAST}} = J(3 + 2M) + M (M+1)/2$. This number, for moderate values of $J$, is much smaller than $v^{\text{VAR}} = M^2P + M (M+1)/2$, the number of parameters of an unrestricted (but linear) VAR.  Notice that the Kronecker structure implies that each equation in the model features the same set of nonlinear transformations of selected covariates. At a first glance, this assumption might be restrictive but if $J$ is set to be large, the model is still flexible enough to capture arbitrary nonlinearities across equations and, specifically, equation-specific idiosyncrasies in terms of nonlinear behavior of the time series.  This is because the corresponding equation-specific parameters for each $\bm Z_t$ can differ. So in case that there is strong evidence that one (or more) variable(s) in the system evolve according to, e.g., a threshold process, our algorithm would add appropriate base learners to the conditional mean model. In this case, the corresponding coefficients would be non-zero whereas the coefficients associated with other transformations would then be close to zero.

In terms of computation, the Kronecker structure in the likelihood  gives rise to substantial computational advantages. This not only relates to posterior sampling (see  \autoref{sec: bayes_inference}) but also to the computation of generalized impulse responses (GIRFs), see \autoref{sec: girfs} of the Online Appendix. GIRF computation in models such as the BART-VAR of \cite{huber2020inference} require computing forecast distributions (both unconditional and conditional on a shock of interest). Since nonlinear models imply that GIRFs are state-dependent, one needs to integrate over the economic conditions. If each equation is determined by its own equation-specific function, this becomes excessively slow and turns out to be the computational bottleneck in these models. The reason is that each equation-specific function, $f_j(\bm x_t)$, needs to be approximated and, for large $M$, the computational burden becomes large. By contrast, our approach only requires us to compute $g(\Tilde{x}_{j, t}|\bm \theta_j)$ for all $j$ and then use the location coefficients to obtain a draw from the conditional mean for $\bm y_t$.  Hence, we do not need to evaluate $g$ for each equation due to the Kronecker structure. And this translates into substantial speed improvements when it comes to computing nonlinear functions such as GIRFs.
\subsection{Bayesian inference}\label{sec: bayes_inference}
Most priors and steps in the posterior simulator remain untouched by moving from the univariate to the multivariate model. Hence, we briefly summarize differences in priors first and then discuss differences to the MCMC algorithm sketched in \autoref{ssec: posterior} 

The priors of the model exactly resemble the ones used for the univariate AST model with two exceptions. We use a Gaussian prior on $\bm \beta$ that conditions on $\bm \Sigma$:
\begin{equation}
    p(\bm \beta | \bm \Sigma) = \mathcal{N}(\bm 0, \bm \Sigma \otimes \underline{\bm V}).
\end{equation}
The prior covariance matrix thus features a Kronecker structure similar to the one in the likelihood. Again, we set $\underline{\bm V} = \phi J^{-1} \times \bm I_{2J}$ and set $\phi = 1$. 

The prior on $\bm \Sigma$ is inverse Wishart:
\begin{equation}
    p(\bm \Sigma) = \mathcal{W}^{-1}(\underline{a}_\Sigma, \underline{\bm S}_\Sigma),
\end{equation}
with $\underline{a}_\Sigma$ denoting prior degrees of freedom and $ \underline{\bm S}_\Sigma$ is a prior scaling matrix. We set $\underline{a}_\Sigma = M$ and $\underline{S}_\Sigma = 1/100 \times \bm I$. This choice yields a proper prior that is relatively uninformative. If one wishes to force a more aggressive model fit, one could set $\underline{\bm S}_\Sigma$ equal to a variance estimator that would imply overfitting and place more weight on the prior by increasing the prior degrees of freedom. 

The posterior simulator differs in three respects from the one associated with the univariate model. First, the particular form of $p(\bm R_j|\nu_j, \mu_j, \bm \delta_j)$ differs, where $\bm R_j$ is now $T \times M$ matrix defined with the $t^{th}$ row given by $(\bm y_t - \sum_{s \neq j} g(\tilde{x}_{s, t}|\bm \theta_s))'$. When we integrate over $\bm \beta$ and $\bm \Sigma$, we end up with the following standard expression for the marginal likelihood of the Bayesian seemingly unrelated regression (SUR) model:
\begin{equation*}
    p(\bm R_j | \nu_j, \mu_j, \bm \delta_j) \propto \left(\frac{|\overline{\bm V}_j|}{|\underline{\bm V}_j|}\right)^{M/2} \times \left(\underline{\bm S}_\Sigma + \bm R'_j \bm R_j + \overline{\bm \beta}_j' \overline{\bm V}^{-1}_j \overline{\bm \beta}_j\right)^{\frac{T + \underline{a}_\Sigma}{2}},
\end{equation*}
where $\overline{\bm \beta}_j$ and $\overline{\bm V}_j$ is defined below \autoref{eq: ML_R}. This expression is used to set up the Metropolis Hastings updates or inverse transform steps employed to sample the thresholds, threshold variables and speed of adjustment parameters.

The next difference relates to how we sample the regression coefficients $\bm \beta$. The full conditional posterior of the multivariate model takes the following form:
\begin{equation}
    p(\bm \beta | \bm \Sigma, \bm Y, \bm Z) = \mathcal{N}(\overline{\bm \beta}, \bm \Sigma \otimes \overline{\bm V}_\beta),
\end{equation}
with $\overline{\bm V}$ being defined as before, $\bm Y = (\bm y_1, \dots, \bm y_T)'$ and $\overline{\bm \beta} = \text{vec}(\overline{\bm V}_\beta \bm Z' \bm Y)$.

Finally, the posterior of $\bm \Sigma$ is inverse Wishart:
\begin{equation*}
    p(\bm \Sigma|\bm Y, \bm Z) = \mathcal{W}^{-1}\left(\underline{a}_\Sigma + T, \underline{\bm S}_\Sigma + \bm Y' \bm Y - \overline{\bm \beta}' \overline{\bm V}^{-1} \overline{\bm \beta}\right). 
\end{equation*}
The resulting MCMC algorithm closely resembles the one discussed in \autoref{ssec: posterior} with the sampling steps for $\bm \beta$, $\bm \Sigma$ and the acceptance/posterior probabilities adjusted accordingly.

It is worth stressing that this algorithm is only slightly more costly than the one for the univariate model. In particular, sampling from the posterior of $\bm \beta$ is more expensive but the Kronecker structure implies that high dimensional matrix operations can be avoided. Hence, sampling from $p(\bm \beta|\bm \Sigma, \bm Y, \bm Z)$ is fast and one can easily estimate nonlinear VARs with more than 100 equations. 

\section{Real-data application} \label{sec: IRFs}
In this section we apply the VAST model to US macroeconomic data. We start by providing a brief overview on the dataset and then move on to provide some evidence on the predictive performance of our model. Finally, we discuss how the US economy reacts to financial shocks.
\subsection{Data overview and model specification}
We apply the VAST model to the FRED-QD dataset \citep{mccracken2020fred}. Our sample runs from $1973$Q1 to $2019$Q4.  In $\bm y_t$, we include $M=80$ variables. These are given in Table \ref{tab:data}. Notice that this set of variables implies that we include a large number of quantities that measure the real side of the economy as well as several factors that capture movements in financial markets. When we consider the effects of financial shocks on the US economy, we also add  the EBP stipulated in \cite{gilchrist2012credit} as a measure of financial conditions. Since this series is only available up to $2016$Q4, we use a slightly shorter sample for the structural analysis. 

In our forecasting exercise we also consider two smaller-sized datasets. These are formed as sub-groups out of this large-scale dataset and defined in Footnote \ref{fn: data}. For the predictive exercise, we drop the EBP to use data through $2019$Q4.

All the models we consider include $p=5$ lags of $\bm y_t$. The number of base learners $J$ is set equal to $50$ when we discuss full sample results (such as the ones in the next sub-section and Sub-section \ref{sec: asymmetries}). We analyze predictive performance over $J$ and find that setting it equal to $40$ or $50$ generally yields the best density forecasting performance. In our structural analysis, we find that changing $J$ leads to impulse responses which are similar in qualitative terms.

\subsection{Predictive evidence}
In this section, we analyze whether our VAST model is capable of outperforming the BART-VAR proposed in \cite{clark2023tail}.\footnote{The setup is precisely the same as the one used in \cite{clark2023tail}.} We include this model because, on a very similar dataset, we have shown that it works well for density and tail forecasts, often improving upon a BVAR with SV and never being substantially outperformed by the BVAR-SV.  To analyze the relationship between model size, density forecasting performance and $J$, we consider three different model sizes and set $J \in \{10, 15, 20, 25, 30, 40, 50\}$. The model sizes we consider are a small-scale (S) model that includes $M=3$ variables. These are the unemployment rate (UNRATE), CPI inflation (CPIAUCSL) and the Federal Funds Rate (FEDFUNDS). The next larger model is a medium-sized (M) one that includes $M=23$ variables. This model uses  the small dataset and adds additional real quantities and financial market variables.\footnote{More precisely, we include the following series from the FRED-QD database: GDPC1, PCECC96, FPIx, GCEC1, INDPRO, CE16OV, UNRATE, CES0600000007, HOUST, PERMIT, PCECTPI, PCEPILFE, GDPCTPI, CPIAUCSL, CPILFESL, CES0600000008, FEDFUNDS, GS1, GS10, M2REAL, TOTRESNS, NONBORRES, S.P.500. The definition of the different abbreviations is given in \autoref{tab:data}. \label{fn: data}} The large-scale (L) dataset is the one described in \autoref{tab:data} bar the EBP and thus $M=79$.

We use a recursive forecasting design that starts with using data through $1989$Q4 to initially train the models. We then compute one-quarter-ahead forecast distributions for $1990$Q1 and evaluate these at the actual outcome using log predictive likelihoods (LPLs).  After obtaining the LPLs for $1990$Q1, we add this data point to the training sample and estimate the one-quarter-ahead predictive density for $1990$Q2 and compute the corresponding LPLs. We repeat this procedure until we reach $2019$Q3 and thus compute the forecasts for $2019$Q4 (the end of the sample). This yields a sequence of time-specific LPLs which we average to end up with the average LPLs we report in \autoref{tab: lpl_forecasts}. This table includes differences between the average LPLs of a particular model and the BART-VAR of a given size.  There are two types of LPLs in the table. One is the marginal LPL for a particular focus variable (UNRATE, CPIAUCSL or FEDFUNDS) whereas the second is the joint LPL for the three focus variables.

\begin{table}[ht!]
\centering
\begin{tabular}{rllrrrrrrr}
  \hline
 &  & $J=$ & 10 & 15 & 20 & 25 & 30 & 40 & 50 \\ 
  \hline
 & UNRATE & S &  0.03 &  0.03 &  0.01 &  0.02 &  0.00 &  0.01 &  0.01 \\ 
   &  & M &  0.02 &  0.03 &  0.05 &  0.05 &  0.05 &  0.01 &  0.00 \\ 
   &  & L & -0.14 &  0.06 &  0.05 &  0.07 &  0.10 &  0.20 &  0.10 \\ \midrule
   & CPIAUCSL & S &  0.08 &  0.11 &  0.12 &  0.14 &  0.12 &  0.13 &  0.15 \\ 
   &  & M & -0.12 & -0.01 & -0.07 & -0.07 &  0.00 & -0.01 & -0.03 \\ 
   &  & L & -0.15 & -0.07 & -0.18 & -0.28 & -0.18 & -0.02 & -0.10 \\ \midrule
   & FEDFUNDS & S & -0.12 & -0.11 & -0.12 & -0.11 & -0.10 & -0.10 & -0.10 \\ 
   &  & M & -0.07 & -0.06 & -0.06 & -0.01 & -0.03 &  0.00 &  0.02 \\ 
   &  & L &  0.05 &  0.07 &  0.10 &  0.09 &  0.13 &  0.15 &  0.18 \\ \midrule
   & Joint & S & -0.03 &  0.02 &  0.03 &  0.05 &  0.03 &  0.05 &  0.05 \\ 
   &  & M & -0.16 & -0.03 & -0.04 & -0.01 &  0.02 & -0.01 & -0.02 \\ 
   &  & L & -0.19 &  0.12 & -0.01 & -0.09 &  0.05 &  0.28 &  0.17 \\ 
   \hline
\end{tabular}
\caption*{\footnotesize \textbf{Notes:} The numbers are the differences between the average log predictive likelihood (based on the one-quarter-ahead density predictions) of the VAST for a specific value of $J$ to the BART-VAR of a particular model size. Averages are computed over the hold-out period (1990Q1 to 2019Q4). S, M and L refer to different model sizes with a precise definition of the included variables given in Footnote XXX.}
\caption{Differences in average one-quarter-ahead density forecasting between the VAST to the BART-VAR across model sizes and for different values of $J$. }\label{tab: lpl_forecasts}
\end{table}

The table indicates that, for specific values of $J$, VAST is capable of improving upon the BART-VAR for almost all target variables and most model sizes. In particular, we find gains that range from being almost zero (such as for small datasets and unemployment forecasts) to modest (such as for inflation arising from small datasets or interest rate forecasts and large datasets). 

For inflation forecasts using medium and large-sized datasets, we find that VAST does not outperform the BART-VAR. But in these cases, setting $J$ either to $40$ or $50$ yields LPLs that are almost identical to the one of the benchmark model. Another case where the BART-VAR produces more accurate density forecasts is the FEDFUNDS rate when the small dataset is adopted.

When we focus on the joint forecasting performance a similar picture arises. We find that VAST yields improvements for small and large models and similar predictive likelihoods for the medium-sized models if $J$ exceeds $10$. For small models, these improvements are muted but consistent across different values of $J$. For the large model, we find that forecast performance varies with $J$ and larger values of $J$ translate into the most precise density forecasts. These joint density forecasts are obtained when we set $J=40$ or $50$. This finding is not surprising given that a larger number of base learners improves flexibility to capture equation-specific nonlinear patterns. 

Next we ask whether the forecasting performance in terms of one-quarter-ahead joint LPLs is heterogenous over time. To do so, we consider the model with $J=40$ as this specification performs well across all focus variables and for joint LPLs  and benchmark it against the large BART-VAR. To understand how performance changes over time, we compute cumulative relative LPLs to the large BART-VAR for VAST across the three different model sizes.

\begin{figure}[h!]
    \centering
    \includegraphics[scale=0.55]{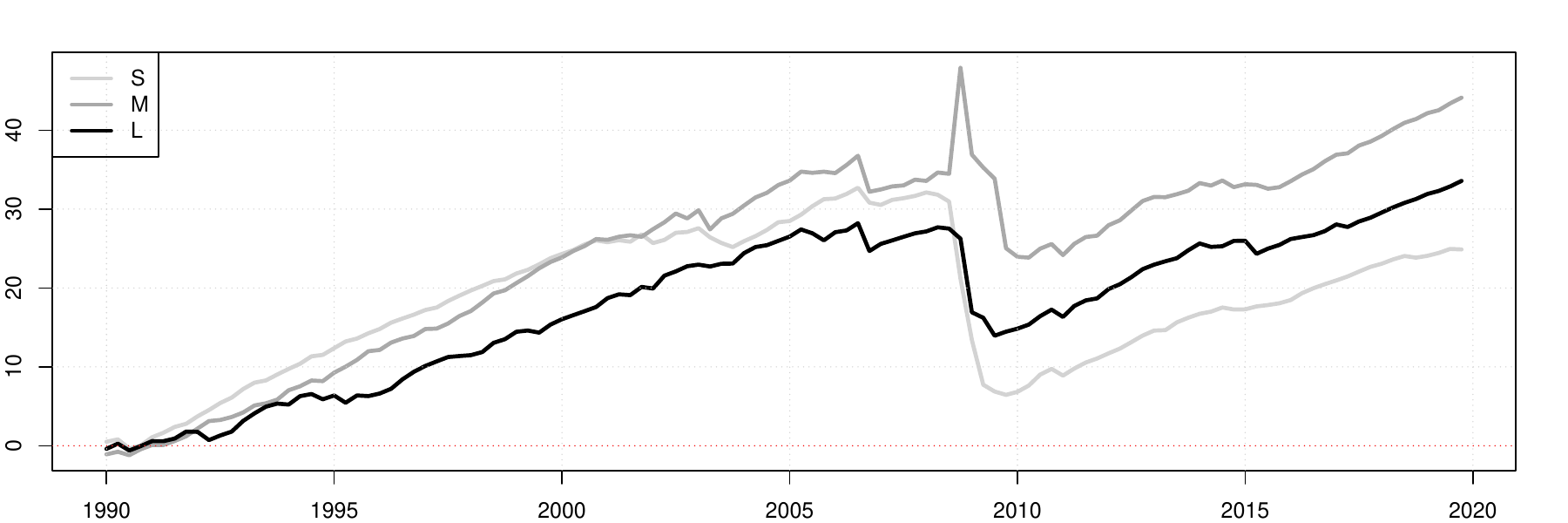}
    \caption{Relative cumulative joint log predictive likelihoods over time relative to the large BART-VAR}
    \label{fig:lpl_ot}
\end{figure}

The results are depicted in \autoref{fig:lpl_ot}. The figure shows that VAST ($J=40$) outperforms the large BART-VAR consistently across all model sizes. Only during the GFC in 2008/2009 we find a slight decline in relative forecasting accuracy for the small and large VAST specifications. In this period, interestingly the medium-scale model outperforms the BART-VAR and relative model performance increases.  Apart from the GFC, the consistent outperformance of VAST remains visible throughout the hold-out period.

This discussion has shown that VAST can improve upon BART, a very competitive benchmark model that has a proven track record in density forecasting. If it is outperformed by the BART-based VAR, the losses in predictive accuracy are typically quite limited. In light of this, it is worth stressing that obtaining the predictive densities of VAST is quick relative to the benchmark. Producing the one-step-ahead density for a particular point in the hold-out takes around five minutes for the large model whereas for the BART-VAR it takes over 1.5 hours on a state-of-the-art Macbook pro.

\subsection{Asymmetric effects of financial shocks}\label{sec: asymmetries}
Next, we turn to the analysis of nonlinearities in the transmission of financial shocks to the US economy. This issue has gained increasing attention in the recent literature \citep[see, e.g.,][]{barnichon2022effects, mumtaz2022impulse, forniJMCB}. Most of these studies find that uncertainty shocks trigger important effects only when they are contractionary and sizable. In all other cases, the effects appear to be muted. This is in stark contrast to the literature utilizing linear VARs \citep{gilchrist2012credit} which find sizable reactions to financial shocks. This is because linear models mix over positive and negative shocks and thus over-exaggerate the effect of a benign shock while underestimating the effect of contractionary shock.

To identify the shock we use zero restrictions that imply that real variables and the Federal Funds rate react sluggishly with respect to a financial shock while financial markets react immediately. This choice is consistent with other papers \citep[see, e.g.,][]{del2020s}that deal with estimating the effects of financial shocks in multivariate time series models.  We use the large dataset with $M=80$ (the $79$ macro series and the EBP) and set $J=50$. 

Our empirical focus will be on two forms of asymmetries. The first is whether benign and adverse shocks trigger different reactions of $\bm y_t$. The second form is whether small shocks trigger different reactions from large shocks. Since our model is nonlinear, asymmetries here could imply that shocks are disproportionally stronger for larger shock sizes or that the shape of the impulse responses differ for small versus large shocks.  We consider a one standard deviation (S.D.) shock  to be a small shock whereas a five S.D. shock is perceived as a large shock.

\begin{figure}[ht!]
    \centering
        \begin{minipage}[t]{.45\textwidth}
        \centering \textbf{(a) Small shock}
        \vspace{.2cm}
    \end{minipage}
    \begin{minipage}[t]{.45\textwidth}
        \centering \textbf{(b) Large shock}
        \vspace{.2cm}
    \end{minipage}\\
    \begin{minipage}[t]{0.45\textwidth}
    \includegraphics[scale=0.4]{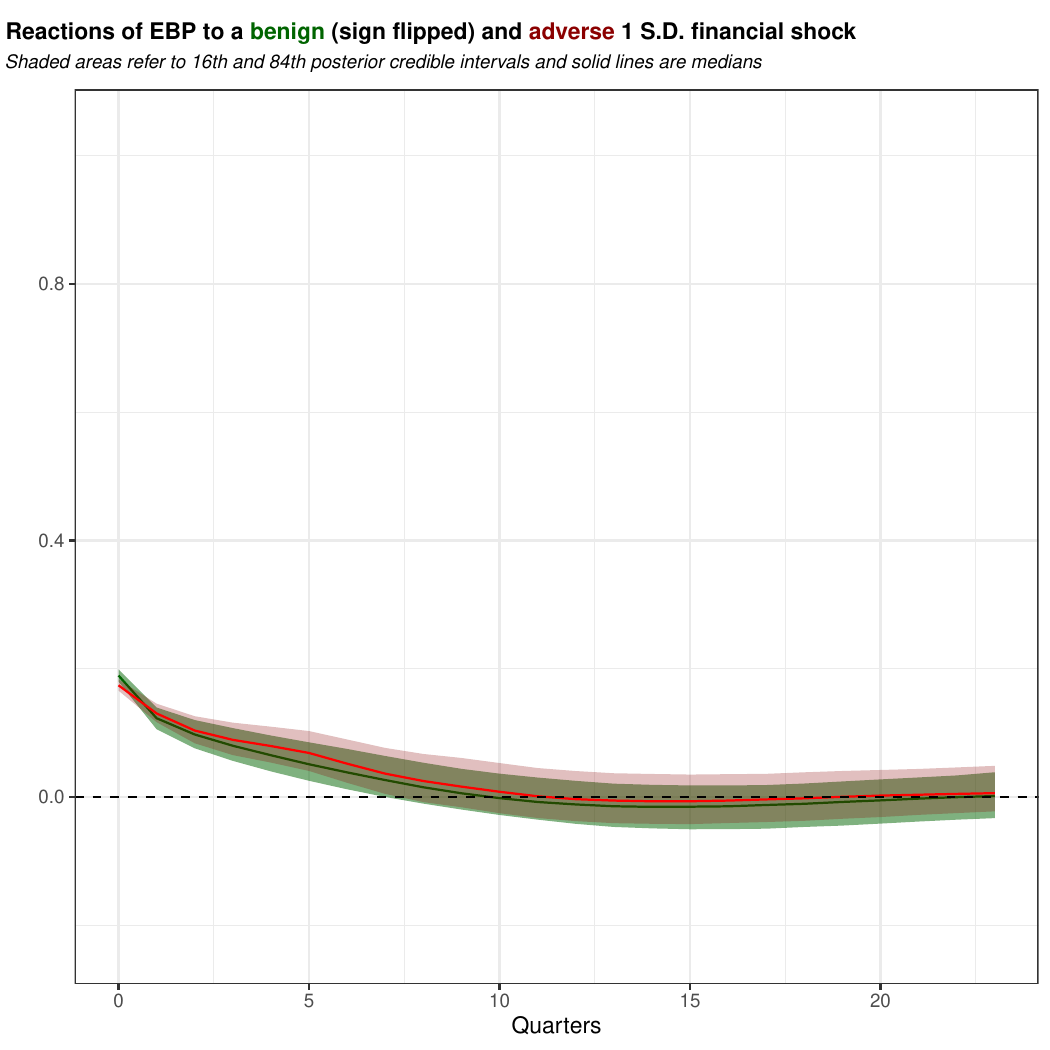}\\
    \end{minipage}
     \begin{minipage}[t]{0.45\textwidth}
    \includegraphics[scale=0.4]{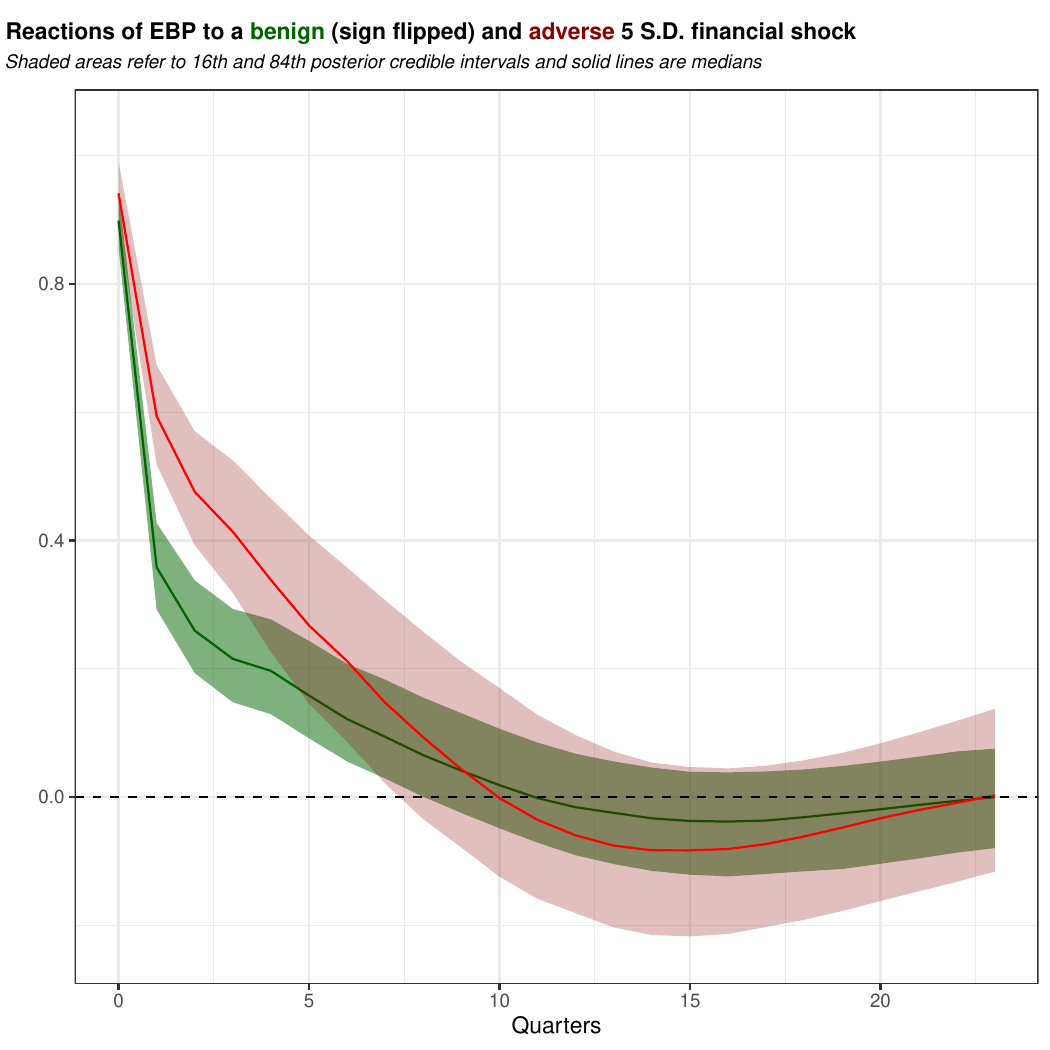}\\
    \end{minipage}
    \caption{Reaction of the Excess Bond Premium to financial shocks}
    \label{fig:EBP_shock}
\end{figure}

We first discuss the endogenous reaction of the EBP to financial shocks. \cite{gilchrist2012credit} argue that fluctuations in the EBP represent movements in investor sentiment or changes in risk preferences in the corporate bond market.  \autoref{fig:EBP_shock}, panels (a) and (b), shows the reaction of the EBP to a small (panel (a)) and a large (panel (b)) financial shock.  

Starting with panel (a) reveals that, if the shock is small, benign and adverse financial shocks trigger a symmetric increase in the EBP which slowly fades out, turning insignificant after around 8 to 10 quarters. By contrast, if the shock size becomes large, we find differences in the shapes of of the EBP reactions. A large and benign financial shock induces a strong immediate reaction that abruptly dies out. An adverse financial shock translates into a strong but more persistent increase in the EBP. Notice that posterior uncertainty is slightly smaller in the case of a benign shock.

\begin{figure}[h!]
    \centering
        \begin{minipage}[t]{.45\textwidth}
        \centering \textbf{(a) Small shock}
        \vspace{.2cm}
    \end{minipage}
    \begin{minipage}[t]{.45\textwidth}
        \centering \textbf{(b) Large shock}
        \vspace{.2cm}
    \end{minipage}\\
    \begin{minipage}[t]{0.45\textwidth}
    \includegraphics[scale=0.4]{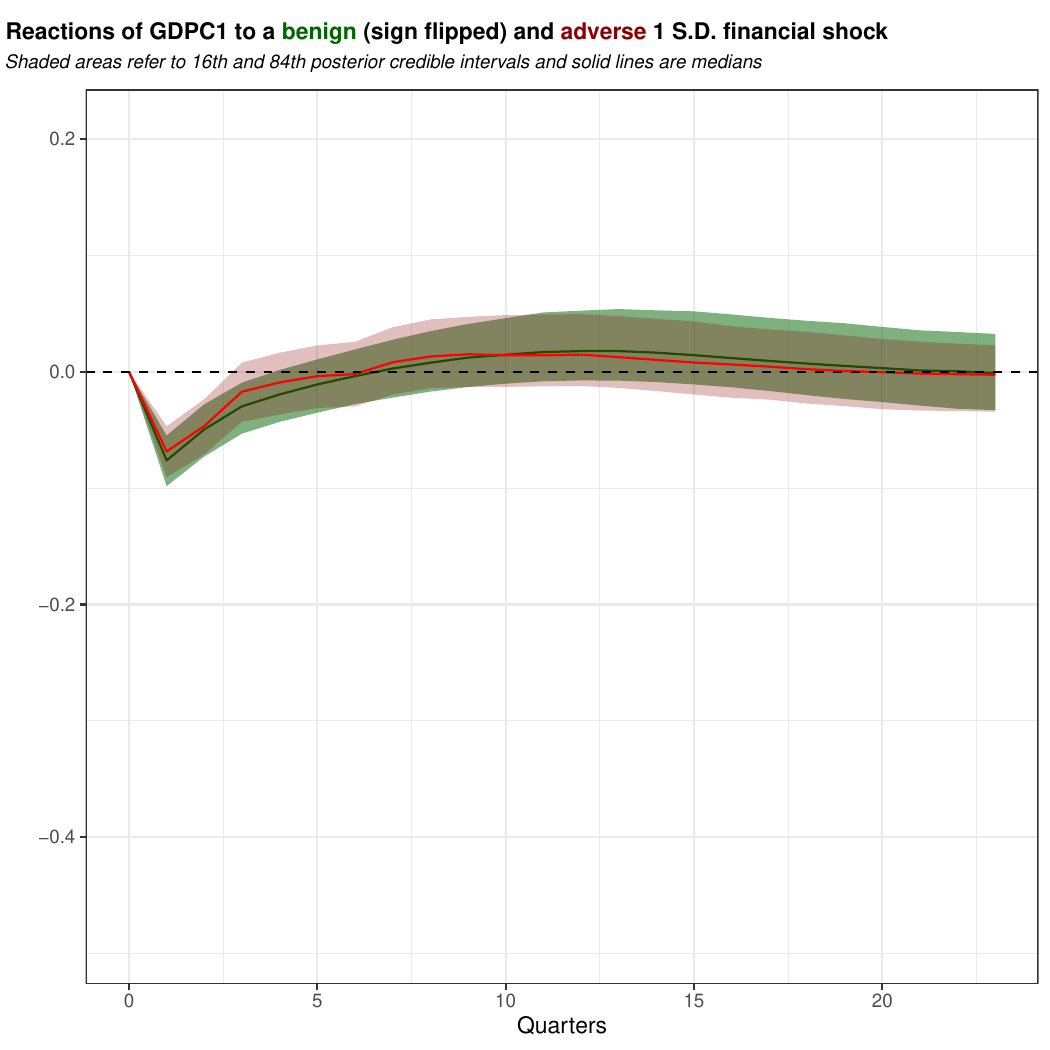}\\
    \end{minipage}
     \begin{minipage}[t]{0.45\textwidth}
    \includegraphics[scale=0.4]{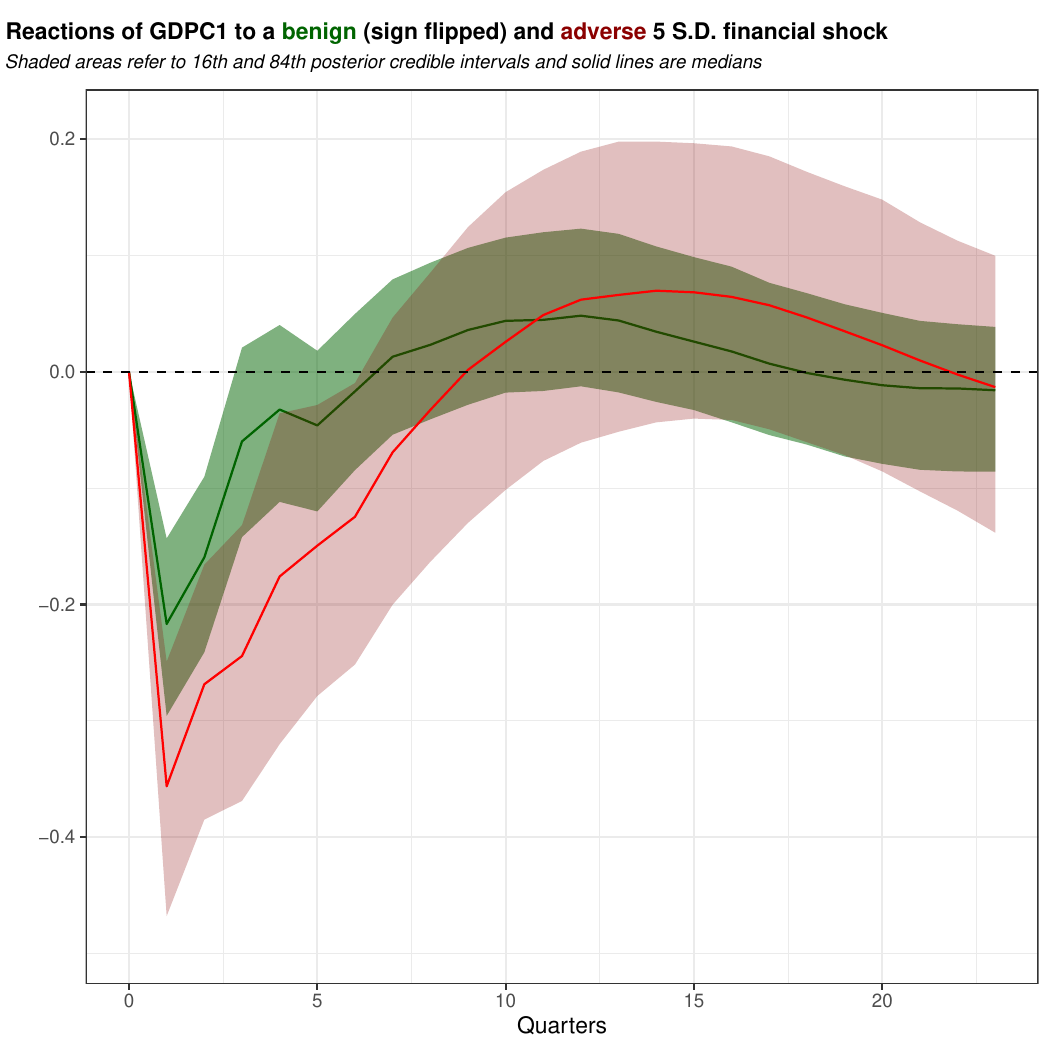}\\
    \end{minipage}\\
        \begin{minipage}[t]{0.45\textwidth}
    \includegraphics[scale=0.4]{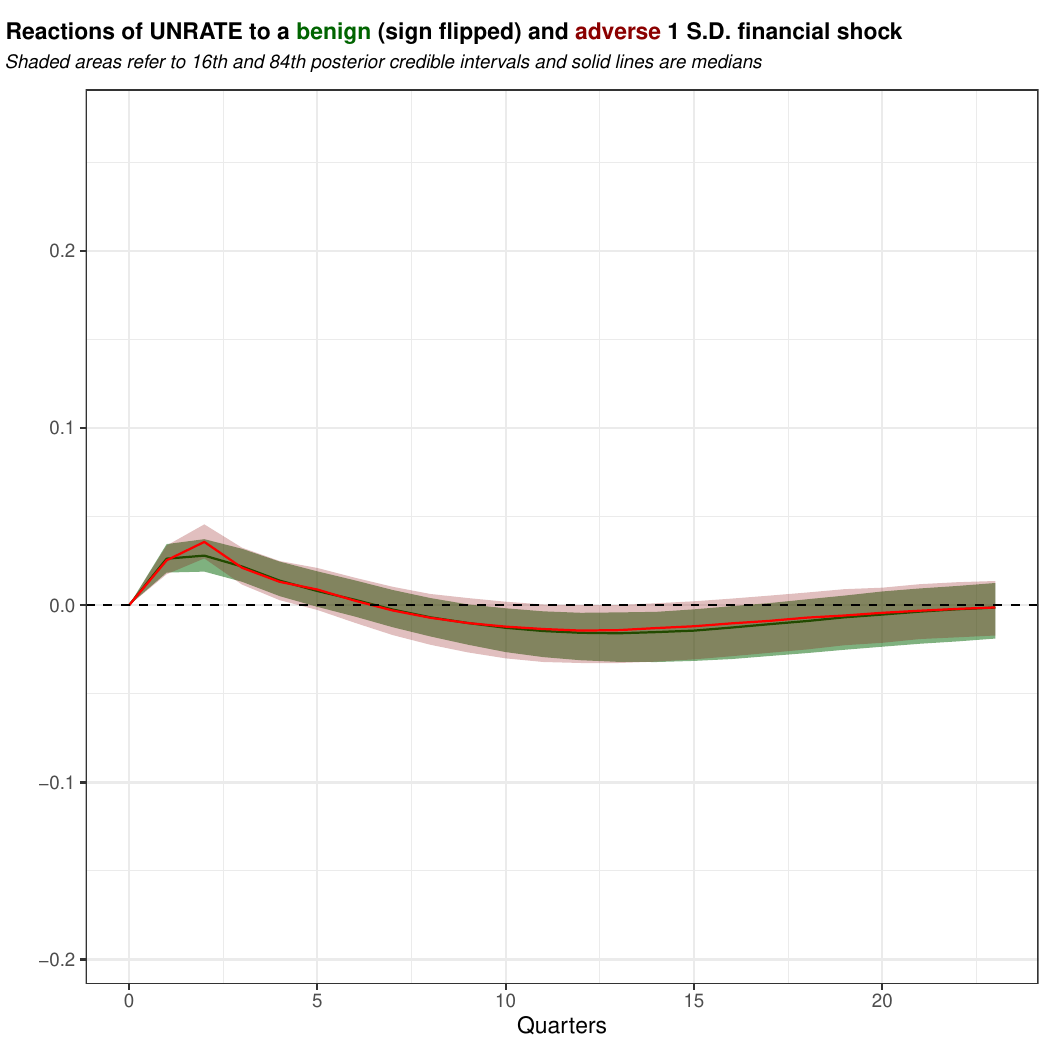}\\
    \end{minipage}
     \begin{minipage}[t]{0.45\textwidth}
    \includegraphics[scale=0.4]{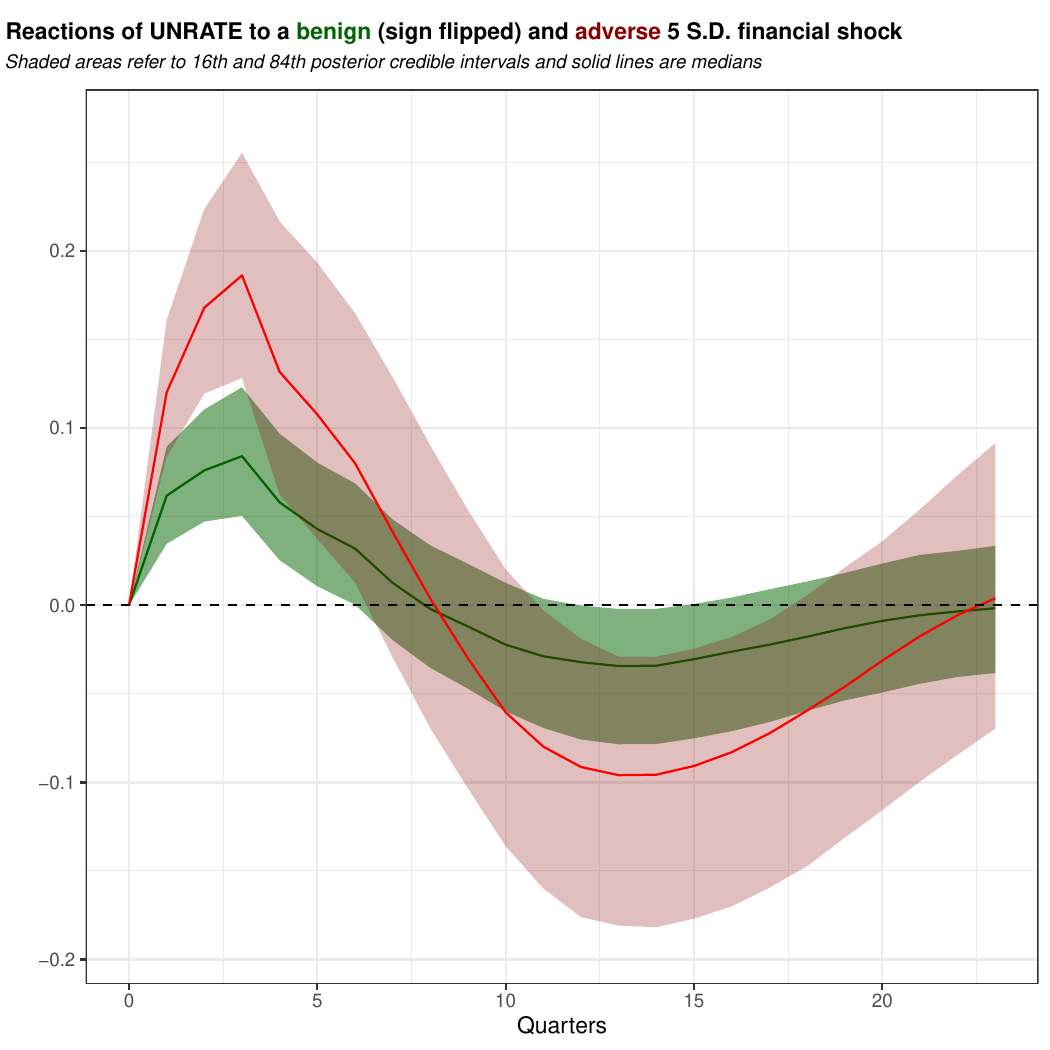}\\
    \end{minipage}
    \caption{Reactions of GDP growth and the unemployment rate to financial shocks}
    \label{fig: gdp_UR_irfs}
\end{figure}
The reaction of the EBP indicates symmetric responses to small shocks (irrespective of sign) but increasing asymmetries if the shock is sizable. This finding carries over to the impulse responses of GDP growth and the unemployment rate in \autoref{fig: gdp_UR_irfs}. In both cases, small financial shocks of either sign trigger reactions of GDP growth and the unemployment rate with the correct sign (i.e. decreasing levels of real activity if the shock is adverse and increasing levels of real activity if the shock is benign). When we consider the effects of large shocks we find substantial asymmetries. In particular, both GDP growth and the unemployment rate display a much stronger reaction to an adverse shock. This is consistent with, e.g., \cite{barnichon2022effects} and \cite{mumtaz2022impulse}, who also document stronger reactions of output growth to contractionary financial shocks. Apart from the stronger short-run effects, we also find that reactions to an adverse shock are slightly more persistent and turn insignificant around two years after the shock hit the system.

\begin{figure}[h!]
   \centering
        \begin{minipage}[t]{.45\textwidth}
        \centering \textbf{(a) Small shock}
        \vspace{.2cm}
    \end{minipage}
    \begin{minipage}[t]{.45\textwidth}
        \centering \textbf{(b) Large shock}
        \vspace{.2cm}
    \end{minipage}\\
    \begin{minipage}[t]{0.45\textwidth}
    \includegraphics[scale=0.4]{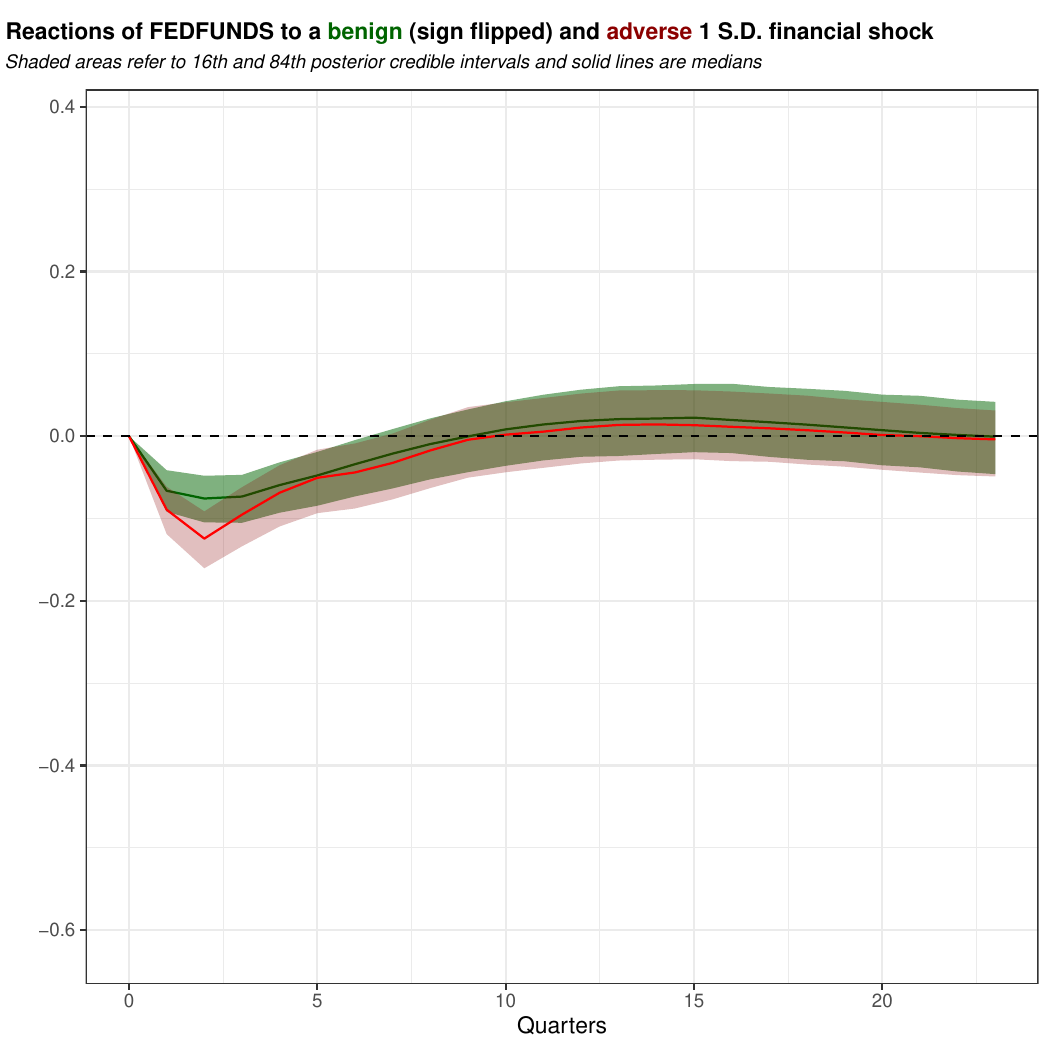}\\
    \end{minipage}
     \begin{minipage}[t]{0.45\textwidth}
    \includegraphics[scale=0.4]{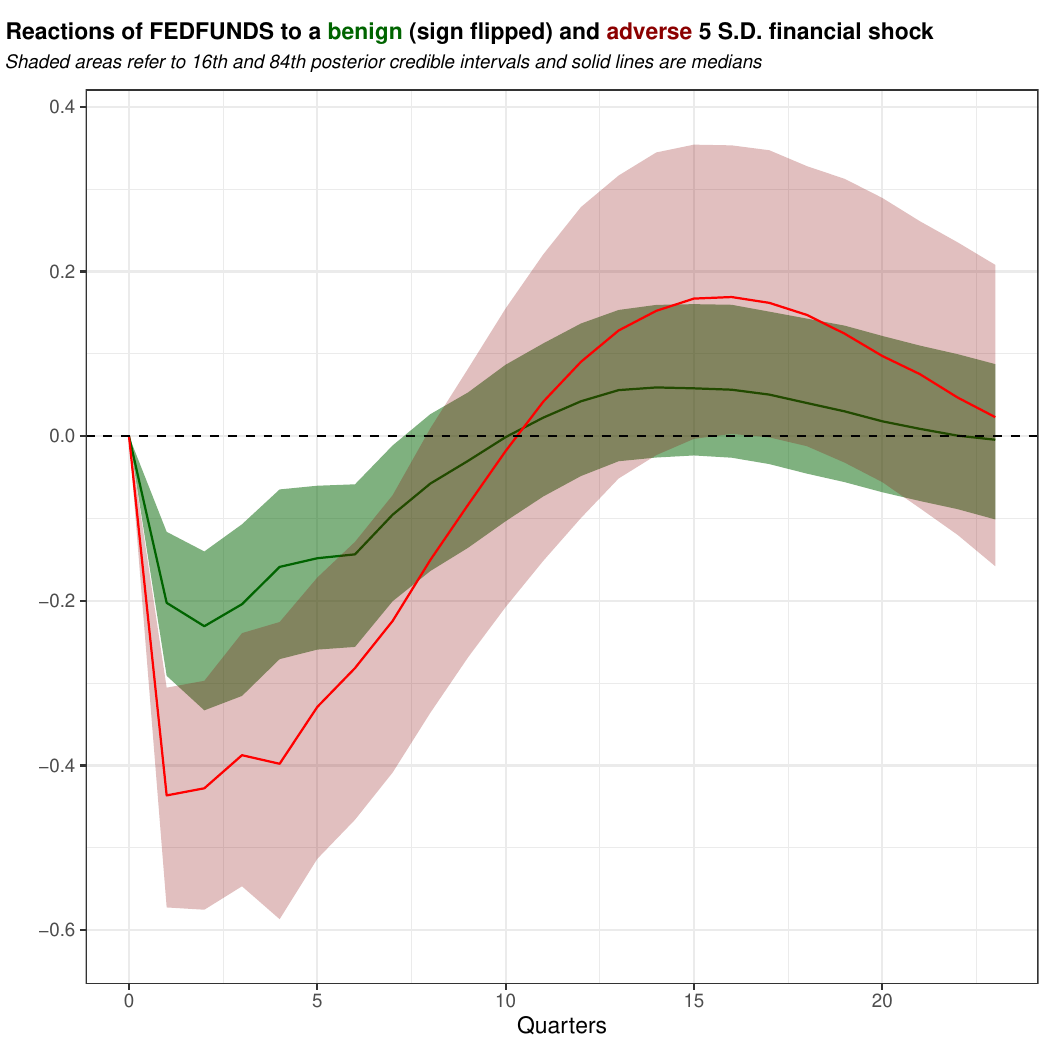}\\
    \end{minipage}\\
        \begin{minipage}[t]{0.45\textwidth}
    \includegraphics[scale=0.4]{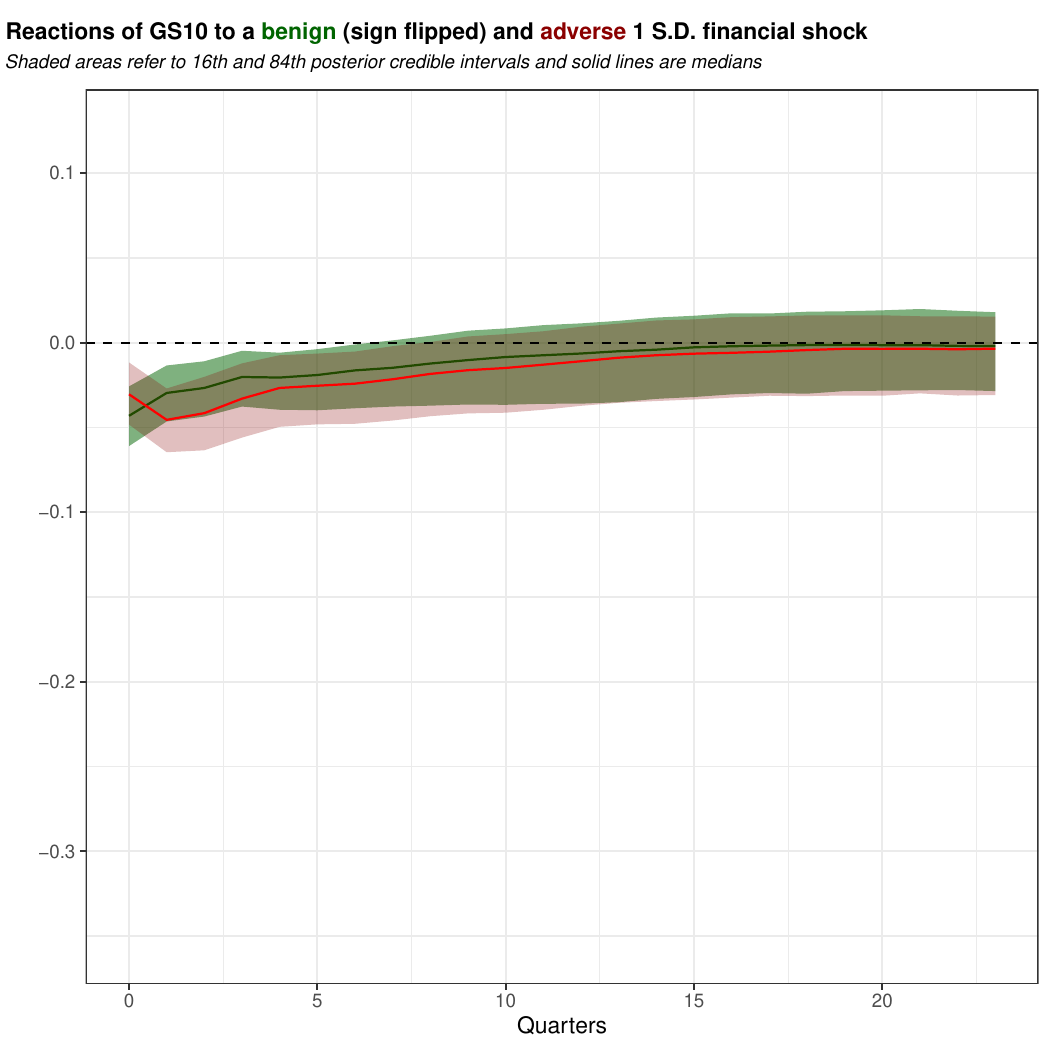}\\
    \end{minipage}
     \begin{minipage}[t]{0.45\textwidth}
    \includegraphics[scale=0.4]{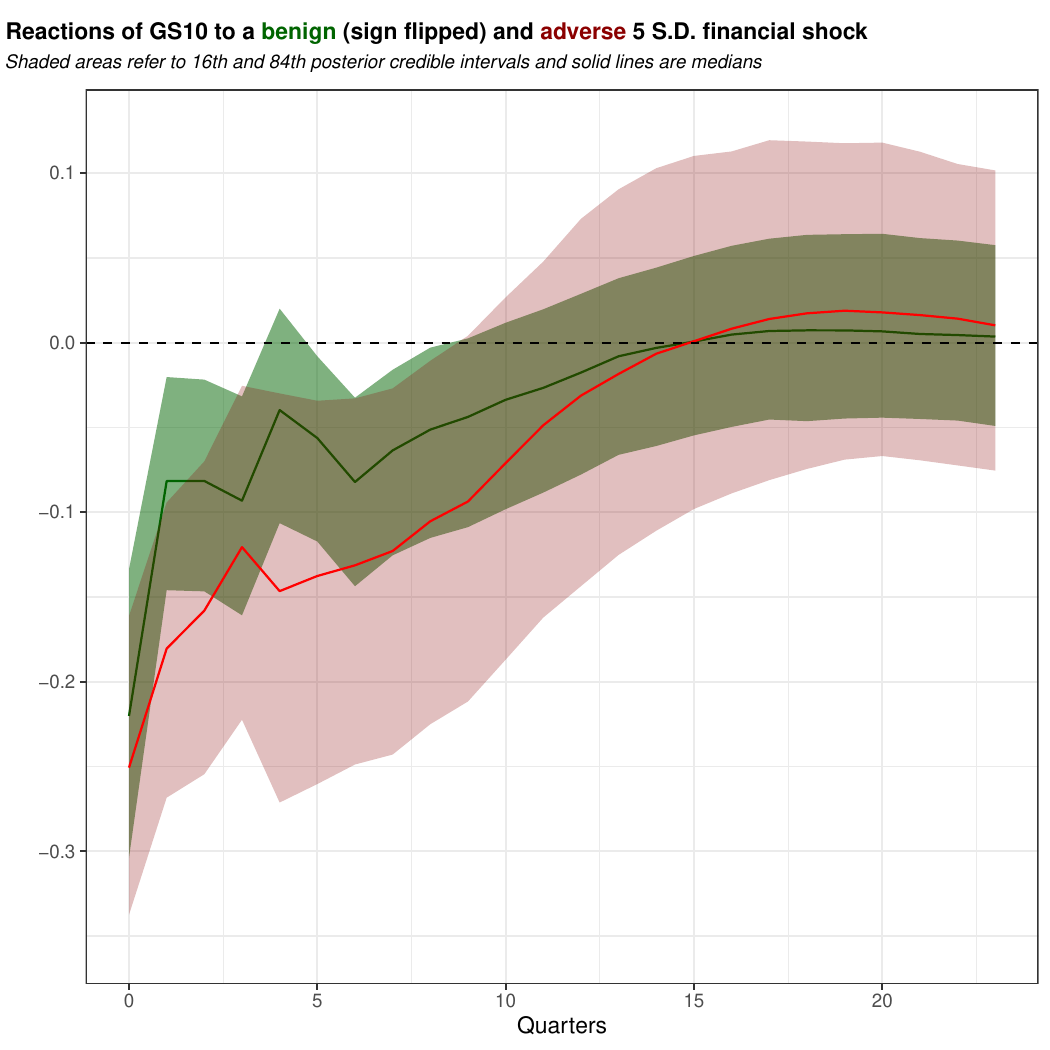}\\
    \end{minipage}
    \caption{Reactions of the federal funds rate and 10-year-yields to financial shocks}
    \label{fig: fedfunds_yields}
\end{figure}

\autoref{fig: fedfunds_yields} shows the responses of short-term interest rates and 10-year US treasury yields. As opposed to output growth and unemployment reactions, the Federal Funds Rate and the 10-year yield reactions feature some asymmetries. For the Federal Funds rate, these asymmetries relate to short-run responses (between one to 1.5 years), with the central bank displaying a stronger reaction in response to a contractionary financial shock. In this case, short-rates are decreased by around 15 basis points (bps) whereas in the benign case, the central bank lowers short-term interest rates by around 10 bps. Treasury yields, by contrast, display a slightly stronger impact reaction to a benign shock but, in the short-run, the decline in response to adverse financial shocks is stronger (in absolute terms) than the increase in yields in response to benign shocks. 

When we consider  large shocks, we find substantial evidence for asymmetries. The Federal Funds rate declines by around 45 bps after one year in response to an adverse shock. The reaction to a benign shock is much more muted, with median increases in short-term interest rates of around 22 bps. For treasury yields, we find similar impact reactions to a large shock (with flipped sign). This is in contrast to a small shock. However, treasury markets exhibit a much stronger reaction to adverse shocks than to benign shocks during the first two years.  

\begin{figure}[ht!]
   \centering
        \begin{minipage}[t]{.45\textwidth}
        \centering \textbf{(a) Small shock}
        \vspace{.2cm}
    \end{minipage}
    \begin{minipage}[t]{.45\textwidth}
        \centering \textbf{(b) Large shock}
        \vspace{.2cm}
    \end{minipage}\\
    \begin{minipage}[t]{0.45\textwidth}
    \includegraphics[scale=0.4]{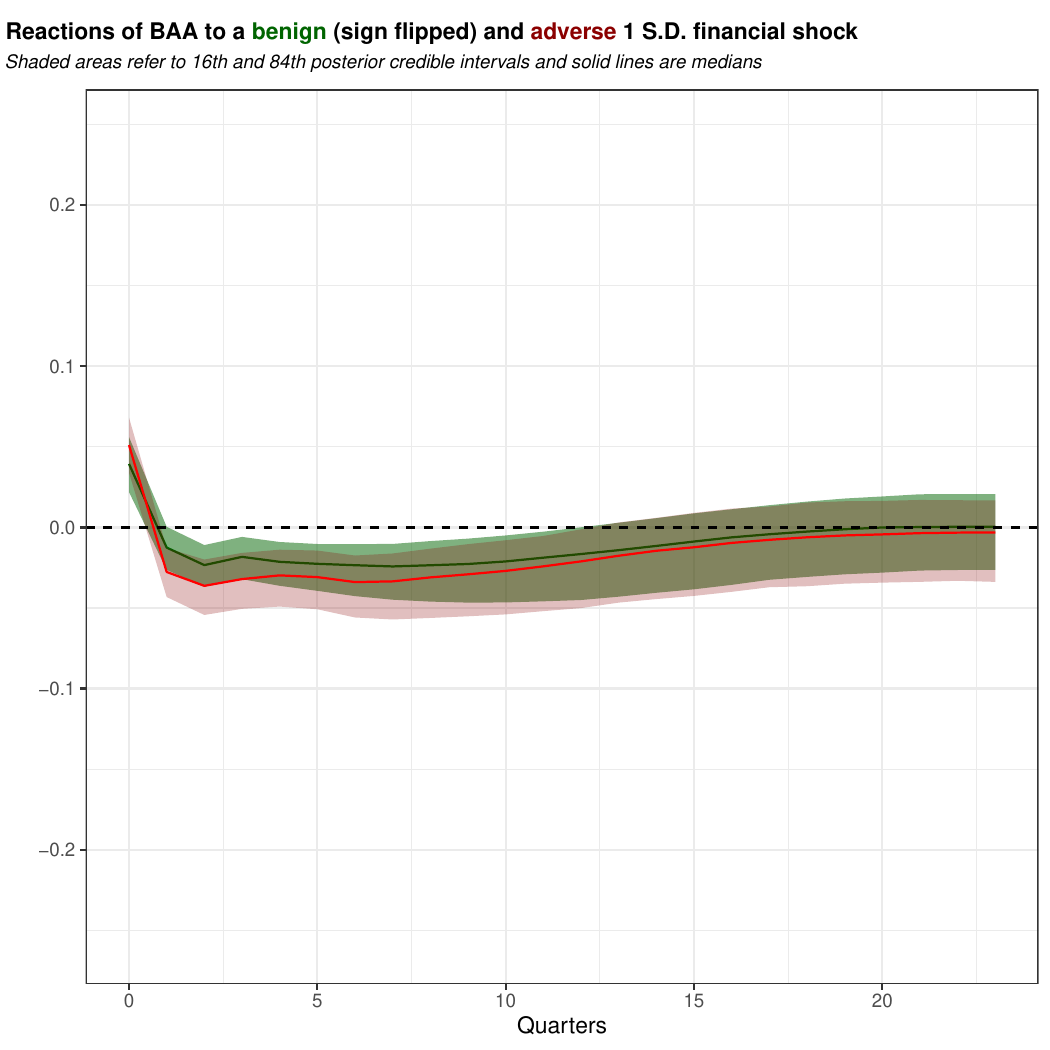}\\
    \end{minipage}
     \begin{minipage}[t]{0.45\textwidth}
    \includegraphics[scale=0.4]{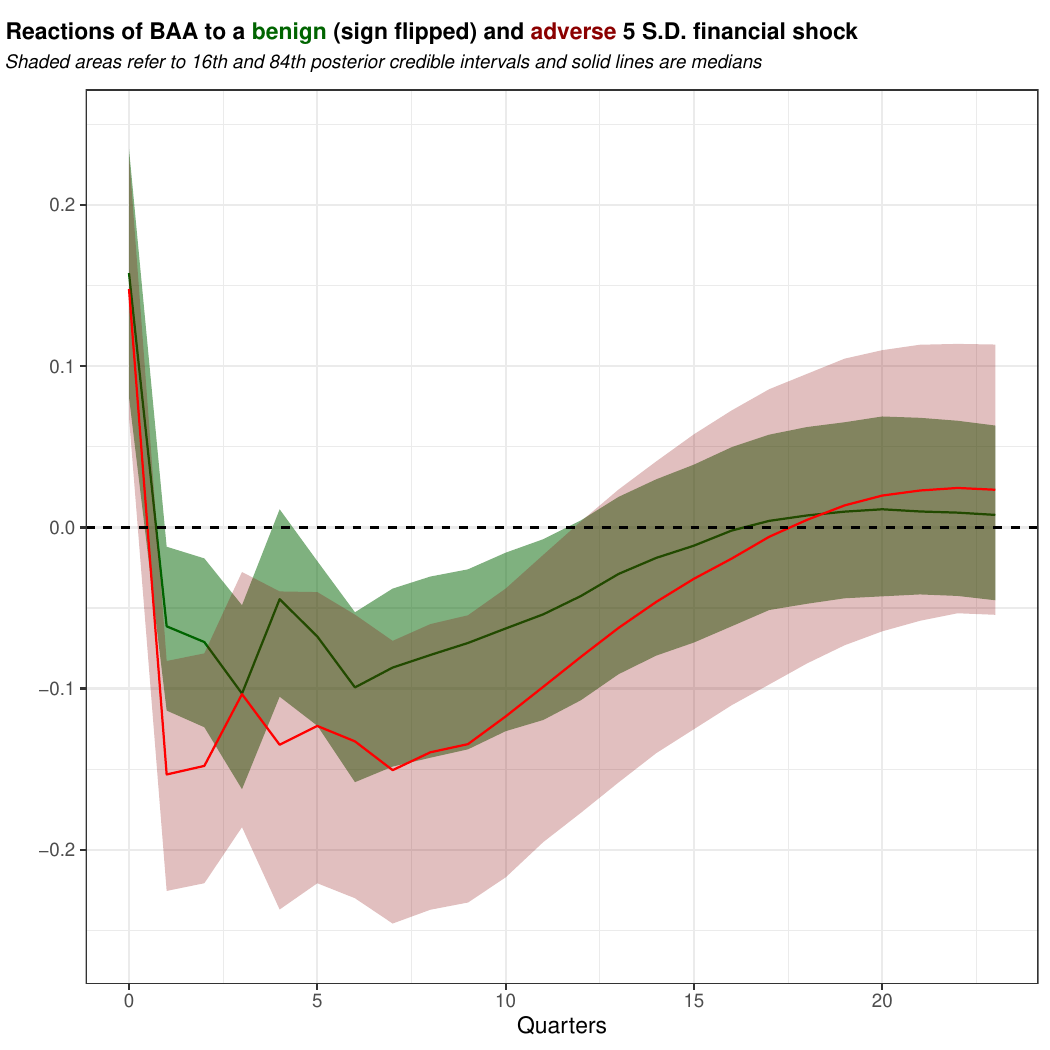}\\
    \end{minipage}\\
        \begin{minipage}[t]{0.45\textwidth}
    \includegraphics[scale=0.4]{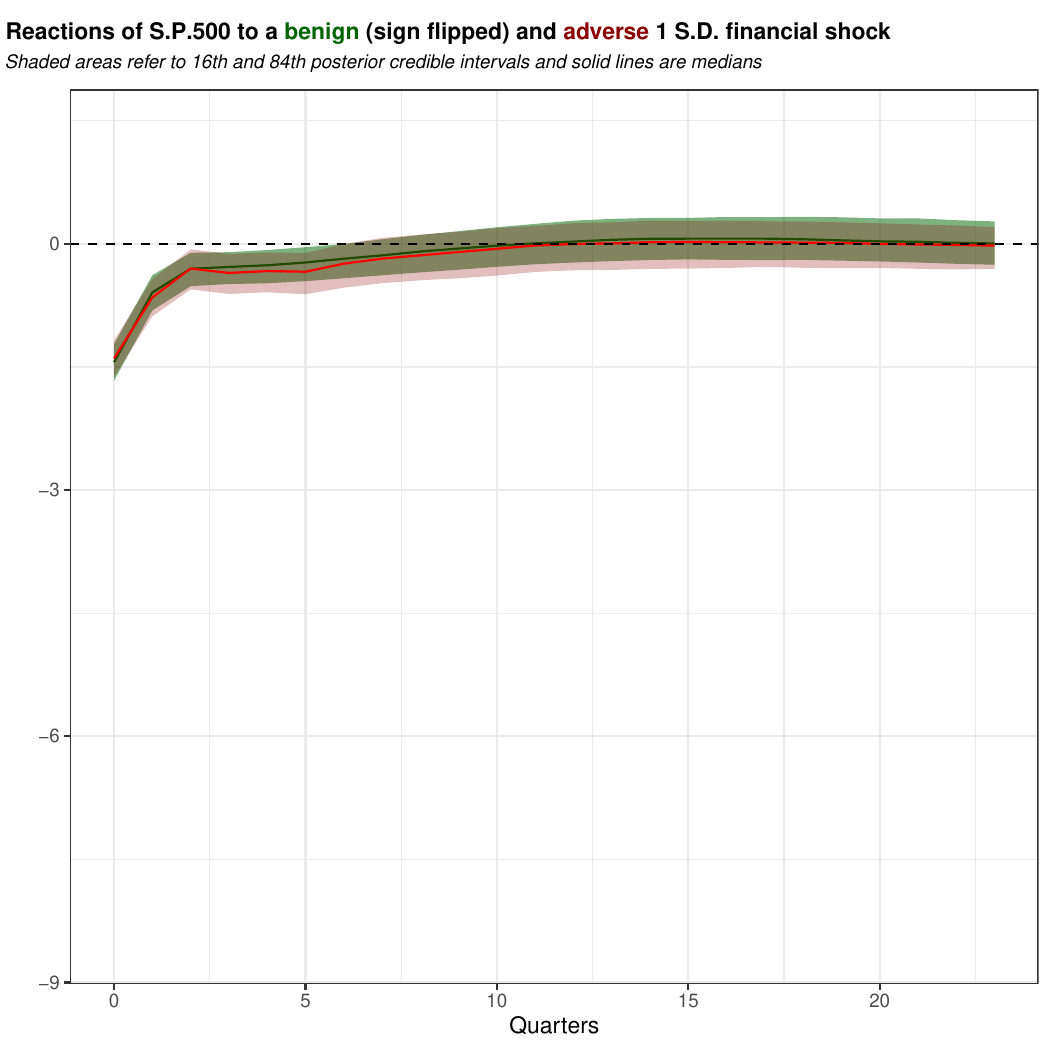}\\
    \end{minipage}
     \begin{minipage}[t]{0.45\textwidth}
    \includegraphics[scale=0.4]{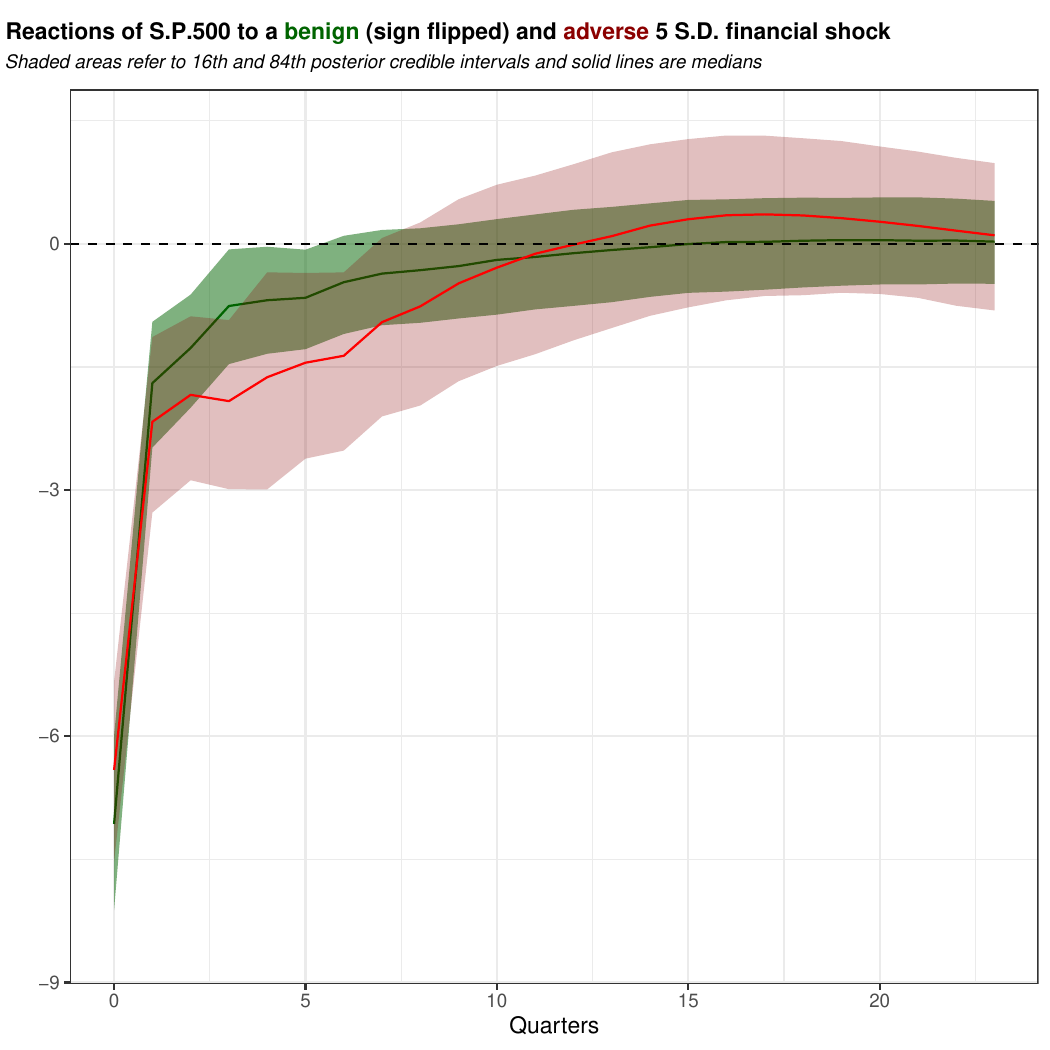}\\
    \end{minipage}
    \caption{Reactions of yields on BAA-rated bonds and the S\&P 500}
    \label{fig:BAAyields_SP500}
\end{figure}

Finally, we consider how other financial markets react to financial shocks. In particular, we focus on the reactions of BAA-rated corporate bond yields and the S\&P 500. Responses of BAA-rated bond yields to small financial shocks point towards no asymmetries for immediate reactions but slightly stronger reactions after around four quarters. Stock markets, by contrast, react symmetrically to small shocks of either sign. When we consider larger shocks we, again, find substantial asymmetries. Reactions to large adverse financial shocks are stronger, in particular between two and 8 quarters. These results, in combination with the reaction of the 10-year treasury yields, can be interpreted in the sense that financial markets are much more reactive to contractionary financial shocks, triggering a 'risk-off' mood of investors. This is reflected by declining stock market returns, increases in BAA-rated bond yields and declines in treasury yields which act as a safe asset.

To sum up, our findings confirm findings in the literature that reactions to adverse shocks are stronger than the ones to benign shocks. However, we also find that this asymmetry result only arises if the shock becomes large. 
\section{Conclusions}
The main goal of this paper is to develop a highly scalable yet flexible econometric model that is capable of capturing asymmetries in possible macroeconomic relations. We achieve this by summing over $J$ simple functions which are location mixtures with transition between regimes driven by a logistic function. Monte Carlo evidence suggests that, in the univariate case,  our approach  produces predictions which are close to the one of BART, a popular machine learning tool, and in some cases slightly more precise. We generalize our approach to the multivariate case, leading to the VAST. This model is highly scalable and can be applied to large datasets.

After showing that our approach works well in predictive terms, we apply the VAST to a dataset with $80$ endogenous variables and consider the asymmetries in the responses to financial shocks. Our results indicate that macro reactions to adverse financial shocks are asymmetric with respect to sign and shock. In particular, we find that positive and negative financial shocks trigger similar reactions if the shock is small. But if the shock becomes large, adverse shocks lead to much stronger reactions.

\small{\setstretch{1.1}
\addcontentsline{toc}{section}{References}
\bibliographystyle{frbc.bst}
\bibliography{References}}\normalsize\clearpage
\setcounter{page}{1} 
\counterwithin{figure}{section}
\counterwithin{table}{section}
\renewcommand\thefigure{\thesection\arabic{figure}}
\renewcommand\thetable{\thesection\arabic{table}}
\begin{appendices}

\setcounter{equation}{0}
 
\renewcommand\theequation{A.\arabic{equation}}
\section{Technical Appendix}
\subsection{Computation of predictive distributions and generalized impulse responses}\label{sec: girfs}
As in many nonlinear models, interpretation of the coefficients is difficult due to the nonlinear transformation of the predictors. The multivariate nature of the VAST makes it even more difficult. For multivariate time series models, researchers are often not interested in specific parameter estimates per se but have a keen interest in how structural shocks affect the dynamics of the observed variables in $\bm y_t$ over time. This is achieved by considering (structural) impulse response functions (IRFs). Another possible way of making use of the VAST model is to employ it to produce forecast distributions for $\bm y_t$. Both, the posterior distribution of the IRFs and the $h=1, \dots, H$-step ahead forecast distributions are not available in closed form and thus additional simulation-based techniques. Moreover, the fact that our model is highly nonlinear calls for generalized impulse responses \citep[GIRFs, see,][]{koop1996impulse} that take this into account. In this sub-section, we first describe how we sample from the $h-$step-ahead forecasts and then how simulation from the posterior of the GIRFs is done.

In general, the $h-$step-ahead predictive distribution is obtained as follows:
\begin{equation}
    p(\bm y_{T+h}|\bm y_T) = \int \int  p(\bm y_{T+h}|\bm y_T, \bm \Sigma, \bm \theta)~ p(\bm \Sigma, \bm \theta|\bm Y)~ d\bm \Sigma d \bm \theta. \label{eq: pred_dens}
\end{equation}
Recall that $\bm \theta = (\bm \beta', \bm \theta_1, \dots, \bm \theta_J)'$ are the parameters associated with the base learners. Unfortunately, this integral can not be solved analytically. Notice that $p(\bm y_{T+h}|\bm y_T, \bm \Sigma, \bm \theta)$ is a conditionally Gaussian component which is obtained iteratively:
\begin{equation*}
    p(\bm y_{T+1}|\bm y_T, \bm \Sigma, \bm \theta) = \mathcal{N}(\overline{\bm y}_{T+1}, \overline{\bm \Sigma}_{T+1}) 
\end{equation*}
with predictive mean and variance given by, respectively:
\begin{align*}
    \overline{\bm y}_{T+1|T} &= \sum_{j=1}^J g(\Tilde{x}_{j, T+1}, \bm \theta_j),\\
    \overline{\bm \Sigma}_{T+1|T} &= \bm \Sigma.
\end{align*}
For two-steps-ahead, we simulate  ${\bm y}^*_{T+1} \sim \mathcal{N}(\overline{\bm y}_{T+1|T}, \overline{\bm \Sigma}_{T+1|T})$ and use ${\bm y}^*_{T+1}$ to set up $\tilde{x}^*_{j, T+2}$ for all $j$. This simulated draw, in turn, is plugged into the conditional mean function again, leading to:
\begin{equation*}
    p(\bm y_{T+2}|\bm y_{T+1}= \hat{\bm y}_{T+1|T}, \bm y_T, \bm \Sigma, \bm \theta) = \mathcal{N}(\overline{ \bm y}_{T+2|T}, \overline{\bm \Sigma}_{T+2|T}).
\end{equation*}
The predictive mean and variance are defined analogously to the one-step-ahead version with $\tilde{x}^*_{j, T+2}$ instead of $\tilde{x}_{j, T+1}$. Repeating this procedure yields the $h$-step-ahead conditional density:
\begin{equation*}
    p(\bm y_{T+h}|\bm y_{T+h-1} = \hat{\bm y}_{T+h-1}, \bm \Sigma, \bm \theta) = \mathcal{N}(\overline{\bm y}_{T+h|T}, \overline{\bm \Sigma}_{T+h|T}).
\end{equation*}
 To sample from the predictive distribution in \autoref{eq: pred_dens}, we sample first sample from $p(\bm \Sigma, \bm \theta|\bm y)$ and then sample from the corresponding Gaussian forecast distribution. The resulting draws are draws from the posterior predictive distribution. Notice that while the distribution $p(\bm y_{T+h}|\bm y_T, \bm \Sigma, \bm \theta)$ is Gaussian, $p(\bm y_{T+h}|\bm y_T)$ can be highly non-Gaussian and feature multiple modes, heavy tails or be skewed. 

 To compute GIRFs we need to discuss how conditional forecasts are produced. The one-step-ahead conditional (on a particular structural shock) predictive density is given by:
 \begin{equation}
     p(\bm y_{t+h}| \bm y_t,  \xi_{j, t} = w),
 \end{equation}
 where $w$ is a real parameter that defines the shock size and sign. This density is obtained similarly to the unconditional one but for $h=0$ we condition on the event that the $j^{th}$ structural shock is set equal to $w$ while the other shocks are obtained from the corresponding marginal distributions. 
 
 Higher-order conditional predictive densities are then obtained as:
 \begin{equation}
     p(\bm y_{t+h}| \bm y_T,  \xi_{j, t} = w, \xi_{j, t+1}=0, \dots, \xi_{j, T+h}=0)
 \end{equation}
 The resulting GIRFs are then obtained by drawing from the the corresponding $h-$step-ahead conditional predictive densities and then based on computing the differences between the draws from the conditional and unconditional predictive distributions.  Since the nonlinear nature of our model implies state dependence we repeat this procedure for all $t$ and then take the mean. By doing so we integrate out the effect of the (observed) states. 


\section{Data Appendix}
\begin{table}[h!]
\centering
\caption{Description of the Dataset}\label{tab:data}
\scalebox{0.55}{
\begin{tabular}{llll}
\toprule
\textbf{Mnemonic} & \textbf{Description} & \textbf{Transformation} & \textbf{Class} \\
\midrule
GDPC1 & Real Gross Domestic Product & 4 & slow \\
PCECC96 & Real Personal Consumption Expenditures & 5 & slow \\
PCESVx & Real Personal Consumption Expenditures: Services & 5 & slow \\
PCNDx & Real Personal Consumption Expenditures: Nondurable Goods & 5 & slow \\
GPDIC1 & Real Gross Private Domestic Investment & 5 & slow \\
FPIx & Real private fixed investment & 5 & slow \\
Y033RC1Q027SBEAx & Real Gross Private Domestic Investment: Fixed Investment: Nonresidential Equipment & 5 & slow \\
PNFIx & Real private fixed investment: Nonresidential & 5 & slow \\
PRFIx & Real private fixed investment: Residential & 5 & slow \\
A014RE1Q156NBEA & Shares of gross domestic product: Gross private domestic investment: Change in private inventories & 1 & slow \\
GCEC1 & Real Government Consumption Expenditures and Gross Investment & 5 & slow \\
EXPGSC1 & Real Exports of Goods and Services & 5 & slow \\
IMPGSC1 & Real Imports of Goods and Services & 5 & slow \\
DPIC96 & Real Disposable Personal Income & 5 & slow \\
INDPRO & IP:Total index Industrial Production Index (Index 2012=100) & 5 & slow \\
IPFINAL & IP:Final products Industrial Production: Final Products (Market Group) (Index 2012=100) & 5 & slow \\
IPCONGD & IP:Consumer goods Industrial Production: Consumer Goods (Index 2012=100) & 5 & slow \\
PAYEMS & Emp:Nonfarm All Employees: Total nonfarm (Thousands of Persons) & 5 & slow \\
CE16OV & Civilian Employment (Thousands of Persons) & 5 & slow \\
UNRATE & Civilian Unemployment Rate (Percent) & 2 & slow \\
UNRATELTx & Unemployment Rate for more than 27 weeks (Percent) & 2 & slow \\
AWHMAN & Average Weekly Hours of Production and Nonsupervisory Employees: Manufacturing (Hours) & 1 & slow \\
AWOTMAN & Average Weekly Overtime Hours of Production and Nonsupervisory Employees: Manufacturing (Hours) & 2 & slow \\
CES0600000007 & Average Weekly Hours of Production and Nonsupervisory Employees: Goods-Producing & 2 & slow \\
CLAIMSx & Initial Claims & 5 & slow \\
HOUST & Housing Starts: Total: New Privately Owned Housing Units Started & 5 & slow \\
PERMIT & New Private Housing Units Authorized by Building Permits & 5 & slow \\
RSAFSx & Real Retail and Food Services Sales (Millions of Chained 2012 Dollars) & 5 & slow \\
PCECTPI & Personal Consumption Expenditures: Chain-type Price Index & 6 & slow \\
PCEPILFE & Personal Consumption Expenditures Excluding Food and Energy & 6 & slow \\
GDPCTPI & Gross Domestic Product: Chain-type Price Index & 6 & slow \\
GPDICTPI & Gross Private Domestic Investment: Chain-type Price Index & 6 & slow \\
IPDBS & Business Sector: Implicit Price Deflator (Index 2012=100) & 6 & slow \\
DGDSRG3Q086SBEA & Personal consumption expenditures: Goods & 6 & slow \\
DDURRG3Q086SBEA & Personal consumption expenditures: Durable goods & 6 & slow \\
DSERRG3Q086SBEA & Personal consumption expenditures: Services & 6 & slow \\
DNDGRG3Q086SBEA & Personal consumption expenditures: Nondurable goods & 6 & slow \\
CPIAUCSL & Consumer Price Index for All Urban Consumers: All Items & 6 & slow \\
CPILFESL & Consumer Price Index for All Urban Consumers: All Items Less Food \& Energy & 6 & slow \\
WPSFD49207 & Producer Price Index by Commodity for Finished Goods & 6 & slow \\
PPIACO & Producer Price Index for All Commodities & 6 & slow \\
WPU0561 & Producer Price Index by Commodity for Fuels and Related Products and Power & 5 & slow \\
OILPRICEx & Real Crude Oil Prices: West Texas Intermediate (WTI) - Cushing, Oklahoma & 5 & slow \\
CPIAPPSL & Consumer Price Index for All Urban Consumers: Apparel & 6 & slow \\
CPITRNSL & Consumer Price Index for All Urban Consumers: Transportation & 6 & slow \\
CPIMEDSL & Consumer Price Index for All Urban Consumers: Medical Care & 6 & slow \\
CUSR0000SAC & Consumer Price Index for All Urban Consumers: Commodities & 6 & slow \\
CES2000000008x & Real Average Hourly Earnings of Production and Nonsupervisory Employees: Construction & 5 & slow \\
CES3000000008x & Real Average Hourly Earnings of Production and Nonsupervisory Employees: Manufacturing & 5 & slow \\
COMPRNFB & Nonfarm Business Sector: Real Compensation Per Hour (Index 2012=100) & 5 & slow \\
CES0600000008 & Average Hourly Earnings of Production and Nonsupervisory Employees: & 6 & slow \\
\bottomrule
\end{tabular}
}
\end{table}
\addtocounter{table}{-1}
\begin{table}
\centering
\scalebox{0.6}{
\begin{tabular}{llll}
\toprule
\textbf{Mnemonic} & \textbf{Description} & \textbf{Transformation} & \textbf{Class} \\
\midrule
FEDFUNDS & Effective Federal Funds Rate (Percent) & 2 & policy \\
EBP & Excess Bond Premium of \cite{gilchrist2012credit} & 1 & fast \\
TB3MS & 3-Month Treasury Bill: Secondary Market Rate (Percent) & 2 & fast \\
TB6MS & 6-Month Treasury Bill: Secondary Market Rate (Percent) & 2 & fast \\
GS1 & 1-Year Treasury Constant Maturity Rate (Percent) & 2 & fast \\
GS10 & 10-Year Treasury Constant Maturity Rate (Percent) & 2 & fast \\
BAA & Moody's Seasoned Baa Corporate Bond Yield (Percent) & 2 & fast \\
TB6M3Mx & 6-Month Treasury Bill Minus 3-Month Treasury Bill, secondary market (Percent) & 1 & fast \\
GS1TB3Mx & 1-Year Treasury Constant Maturity Minus 3-Month Treasury Bill, secondary market & 1 & fast \\
GS10TB3Mx & 10-Year Treasury Constant Maturity Minus 3-Month Treasury Bill, secondary market & 1 & fast \\
CPF3MTB3Mx & 3-Month Commercial Paper Minus 3-Month Treasury Bill, secondary market & 1 & fast \\
GS5 & 5-Year Treasury Constant Maturity Rate & 2 & fast \\
TB3SMFFM & 3-Month Treasury Constant Maturity Minus Federal Funds Rate & 1 & fast \\
T5YFFM & 5-Year Treasury Constant Maturity Minus Federal Funds Rate & 1 & fast \\
AAAFFM & Moody's Seasoned Aaa Corporate Bond Minus Federal Funds Rate & 1 & fast \\
M1REAL & Real M1 Money Stock & 5 & fast \\
M2REAL & Real M2 Money Stock & 5 & fast \\
BUSLOANSx & Real Commercial and Industrial Loans, All Commercial Banks & 5 & fast \\
CONSUMERx & Real Consumer Loans at All Commercial Banks & 5 & fast \\
NONREVSLx & Total Real Nonrevolving Credit Owned and Securitized, Outstanding & 5 & fast \\
REALLNx & Real Real Estate Loans, All Commercial Banks & 5 & fast \\
TOTALSLx & Total Consumer Credit Outstanding & 5 & fast \\
TOTRESNS & Total Reserves of Depository Institutions & 6 & fast \\
NONBORRES & Reserves Of Depository Institutions, Nonborrowed & 7 & fast \\
EXSZUSx & Switzerland / U.S. Foreign Exchange Rate & 5 & fast \\
EXJPUSx & Japan / U.S. Foreign Exchange Rate & 5 & fast \\
EXUSUKx & U.S. / U.K. Foreign Exchange Rate & 5 & fast \\
EXCAUSx & Canada / U.S. Foreign Exchange Rate & 5 & fast \\
S.P.500 & S\&P's Common Stock Price Index: Composite & 4 & fast \\
\bottomrule
\end{tabular}
}
 \smallskip
\begin{minipage}{\linewidth}\small
\tiny \textbf{Notes}: This table provides an overview of the dataset employed. The transformation codes are applied to each time series  and described in \cite{mccracken2020fred}: (1) no transformation; (2) $\Delta y_{jt}$; (3) $\Delta^2 y_{jt}$; (4) $\log (y_{jt})$; (5) $\Delta \log (y_{jt})$; (6) $\Delta^2 \log (y_{jt})$; (7) $\Delta (y_{jt}/y_{jt-1} - 1)$. The column 'Class' indicates whether a variable is fast or slow moving.
\end{minipage}

\end{table}
\end{appendices}
\end{document}